\documentclass[12pt]{iopart}
\expandafter\let\csname equation*\endcsname\relax
\expandafter\let\csname endequation*\endcsname\relax
\usepackage{amsmath}
\usepackage[pdftex]{graphicx}
\setkeys{Gin}{clip=true,draft=false}
\graphicspath{{./pdf/}}
\DeclareGraphicsExtensions{.pdf}

\usepackage{soul,color}

\makeatletter
\def\vec@style{\relax} 
\def\vec#1{\relax\ifmmode\mathchoice
{\mbox{\boldmath$\vec@style\displaystyle#1$}}
{\mbox{\boldmath$\vec@style\textstyle#1$}}
{\mbox{\boldmath$\vec@style\scriptstyle#1$}}
{\mbox{\boldmath$\vec@style\scriptscriptstyle#1$}}\else
\hbox{\boldmath$\vec@style\textstyle#1$}\fi}

\def\mat@style{\sf} 

\def\mat#1{\relax\ifmmode\mathchoice
{\mbox{\boldmath$\mat@style\displaystyle#1$}}
{\mbox{\boldmath$\mat@style\textstyle#1$}}
{\mbox{\boldmath$\mat@style\scriptstyle#1$}}
{\mbox{\boldmath$\mat@style\scriptscriptstyle#1$}}\else
\hbox{\boldmath$\mat@style\textstyle#1$}\fi}
\makeatother

\begin{document}
\title{Memcomputing with membrane memcapacitive systems}
\author{Y. V. Pershin$^1$, F. L. Traversa$^2$, M. Di Ventra$^2$}
\address{$^1$ Department of Physics and Astronomy, University of South Carolina, Columbia, South Carolina 29208, USA}
\address{$^2$ Department of Physics, University of California, San Diego, La Jolla, California 92093-0319, USA}
\eads{\mailto{pershin@physics.sc.edu}, \mailto{ftraversa@physics.ucsd.edu}, \mailto{diventra@physics.ucsd.edu}}


\begin{abstract}
We show theoretically that networks of membrane memcapacitive systems -- capacitors with memory made out of membrane materials -- can be used to perform a complete set of logic gates in a massively parallel way by simply changing the external input amplitudes, but not the topology of the network. This {\it polymorphism} is an important characteristic of memcomputing (computing with memories) that closely reproduces one of the main features of the brain. A practical realization of these membrane memcapacitive systems, using, e.g., graphene or other 2D materials, would be a step forward towards a solid-state realization of memcomputing with passive devices.
\end{abstract}
\maketitle

\section{Introduction}
Memcomputing, namely computing {\it with} and {\it in} memory, is a novel non-Turing paradigm of computation that employs memory elements to process and store information at the same physical location \cite{diventra13a,traversa14b}. Even though this paradigm could be realized with standard complementary metal-oxide-semiconductor (CMOS) technology \cite{pershin10c}, its main premises rest on the use of passive circuit elements with memory (memelements), namely, memristive,~ \cite{chua76a} memcapacitive and meminductive systems~ \cite{diventra09a}. These memelements can indeed
find numerous applications in electronics, including bio-inspired circuits, \cite{pershin09b,traversa13a} neuromorphic circuits \cite{jo10a,Kim12,pershin12a} and various unconventional computing architectures \cite{diventra13a,borghetti10a,pershin10c,pershin11d} -- just to name a few.

Memcomputing can be employed in both analog and digital mode \cite{diventra13a,traversa14b} also combined with standard CMOS technology. \cite{Kim12,xia09a} The first mode of operation is ideal for the solution of optimization problems, otherwise difficult to solve using standard digital machines \cite{traversa14b,pershin11d,pershin13b}, as well as for analog computing \cite{Wright11a}. Moreover, memristive \cite{pershin10c,thomas2013memristor} and memcapacitive \cite{pershin14a} neural networks can also be considered as an analog realization of memcomputing. The second -- digital -- mode combines the strengths of memcomputing (most notably its intrinsic massive parallelism) with standard digital logic functionality.

The possibility of performing logic operations directly in memory with memelements \cite{borghetti10a,linn2012beyond,diventra13a,traversa14b,ievlev14a} could also solve the long-standing von Neumann bottleneck problem \cite{Backus78a} of modern computer architectures. Since usual capacitors have practically very low dissipations, memcapacitive systems are ideal components to perform computation with little energy \cite{traversa14a}, thus offering a solution to another pressing problem in modern computers: the ever-increasing energy consumption of our digital machines \cite{Kogge11a}.

In previous work we have suggested the use of solid-state memcapacitive systems with diverging and negative capacitance \cite{martinez09a} to perform logic operations within an architecture inspired by the dynamic random access memory one. We called this architecture a Dynamic Computing Random Access Memory (DCRAM) \cite{traversa14a}. It is worth noting, however, that although previously used solid-state memcapacitive systems can be used in digital mode, they are intrinsically analog elements. Memcapacitive systems that are fundamentally digital would thus be a better fit for this type of application although it is difficult to achieve 3D integration with membrane memcapacitors.

\begin{figure}[h]
 \begin{center}
  \includegraphics[width=12cm]{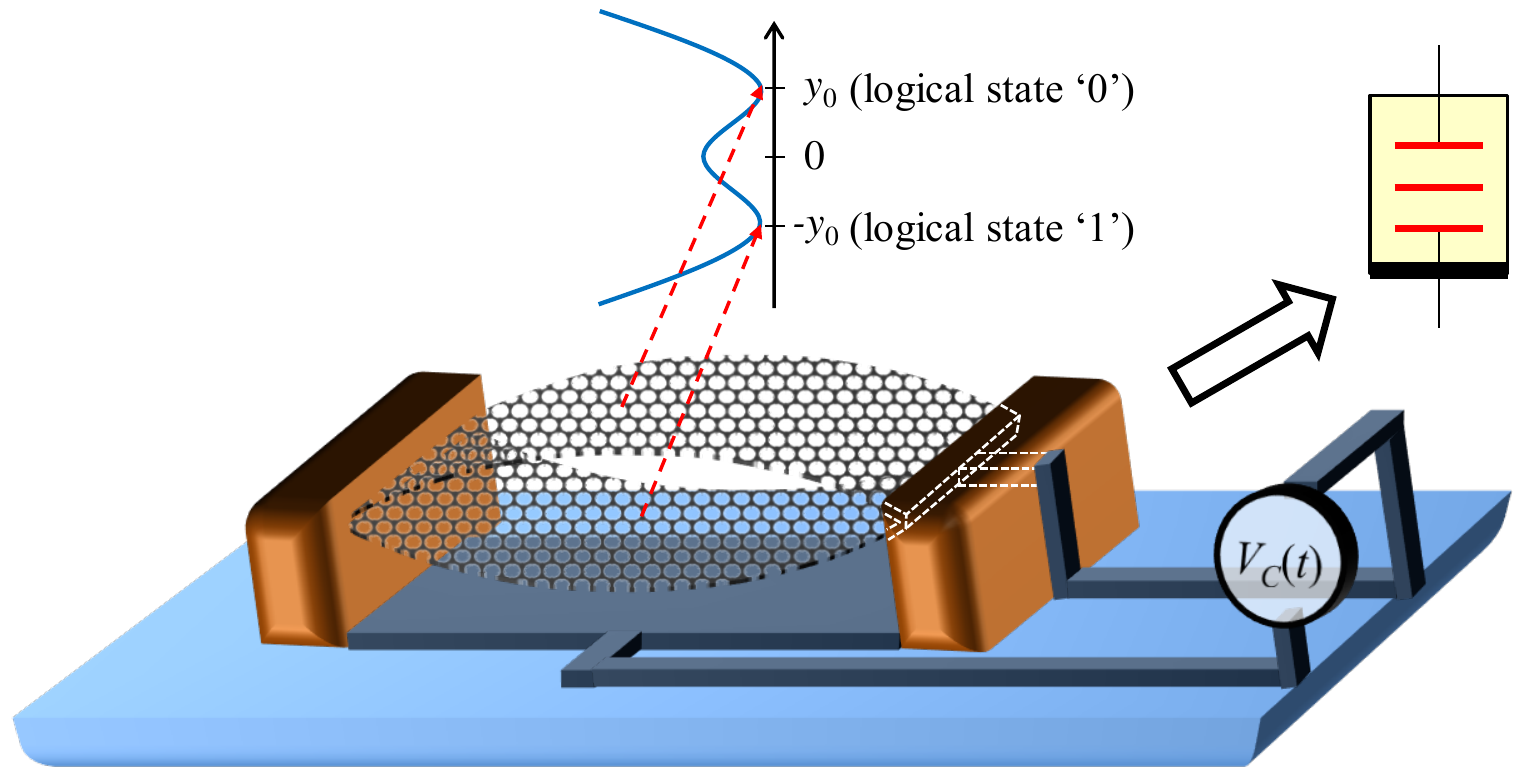}
\caption{\label{fig1} Schematics of membrane memcapacitive system and double well potential describing two equilibrium positions of the membrane at zero voltage.}
\end{center}
\end{figure}

In this paper we employ the class of membrane memcapacitive systems \cite{pershin11c} in the area of binary computing. Membrane memcapacitive systems (see a schematic in Fig.~\ref{fig1}(a)) fit ideally in this context since stressed membranes have only two stable states unlike analog realizations of memcapacitive systems \cite{Liu06a,Lai09a,martinez09a,Flak14a}. The energy barrier between these two states plays the role of an intrinsic threshold that automatically assigns a binary value to any intermediate final state of the system. The combination of many of these membrane memcapacitive elements in an architecture like the DCRAM we have previously analyzed \cite{traversa14a} would then represent an alternative way of implementing memcomputing in the solid state with passive devices. It is important to notice that such systems can be realized using graphene membranes~ \cite{eichler11,stoller08,ElKadi13,Singh14a} or any other molecular system as the flexible plate, and therefore our predictions are within reach of experimental verification.

There are several potential advantages of logic circuits based on membrane memcapacitive systems compared to traditional logic architectures. As mentioned above, the information processing and storage occur on the same physical platform. This feature reduces the amount of information transfer inside the computing system, bypassing altogether the data transfer between the memory and the central processing unit (CPU) where information is traditionally processed. In addition, it allows performing logic operations in memory in a massively-parallel and polymorphic way \cite{traversa14a}. This latter feature means that the implementation of different logic functions is not based on any specific pre-wired structure. Rather, the type of logic operation is selected only by the control signal amplitudes. As such, memcapacitive logic circuits require much smaller number of individual components as well as offer a versatility inexistent in traditional logic circuits, even those employing memristive components \cite{borghetti10a}.


\section{Membrane memcapacitive system} 

By definition \cite{diventra09a}, a voltage-controlled memcapacitive system is given by the equations
\begin{eqnarray}
q(t)&=&C\left(x,V_C,t \right)V_C(t) \label{Ceq1} \\
\dot{x}&=&f\left( x,V_C,t\right) \label{Ceq2}
\end{eqnarray}
where $q(t)$ is the charge on the capacitor at time $t$, $V_C(t)$ is the applied voltage, $C$ is the {\it memcapacitance}, $x$ is a set of $n$ state variables describing the internal state of the system, and $f$ is a continuous $n$-dimensional vector function. It is important that the memcapacitance $C$ depends on the state of the system and can vary in time. Some theoretical and experimental studies of memcapacitive effects can be found in the literature \cite{Liu06a,martinez09a,Lai09a,Biolek10b,krems2010a,pershin11c,Flak14a} (for a recent review, see Ref. \cite{pershin11a}).

\subsection{Membrane memcapacitor model} \label{sec21}
In the membrane memcapacitive system \cite{pershin11c}, the mechanism of memory capacitance is geometrical \cite{pershin11a}. In this structure, the capacitor is formed by a strained membrane (upper plate) and a flat fixed lower plate as shown in Fig.~\ref{fig1}. Two equilibrium states of flexible membrane (up-bent and down-bent) are suitable for non-volatile storage of bits of information. When the membrane is in a position closer to the bottom plate, the capacitance of the device is higher -- we call this configuration '1'. When the membrane is bent up, the system has lower capacitance denoted by '0'.

A mathematical model of the membrane memcapacitive system formulated in Ref. \cite{pershin11c} is based on a double-well potential (Fig.~\ref{fig1}). This model describes the bistable membrane device as a second-order voltage-controlled memcapacitive system \cite{diventra09a} in terms of the following equations:
\begin{eqnarray}
q(t)&=& C(y) V_C(t), \label{eqC1} \\
\frac{\textnormal{d}y}{\textnormal{d}\tau}&=&\dot{y}, \label{eqC2}\\
\frac{\textnormal{d}\dot{y}}{\textnormal{d}\tau}&=&
-4\pi^2\,y\,\left(\left(\frac{y}{y_0}\right)^2-1\right)-\Gamma\,\dot{y}-\left(\frac{\beta(\tau)}{1+y}\right)^2, \;\;\;\; \label{eqC3}
\end{eqnarray}
where
\begin{equation}
C(y)=\frac{C_0}{1+y},
\end{equation}
$y=z/d$, $z$ is the position of the top membrane with respect to its middle position, $d$ is the separation between the bottom plate and middle position of the flexible membrane, $y_0=z_0/d$, $\Gamma=2\pi\,\gamma/\omega_0$, $\beta(t)=2\pi / \left( \omega_0\,d \right) \sqrt{C_0 / \left( 2\,m\right)}\,V_C(t)$ and time derivatives are taken with respect to the dimensionless time $\tau=t\,\omega_0 / \left( 2\pi \right)$. Here, $\pm z_0$ are the equilibrium positions of the membrane, $\gamma$ is the damping constant, $\omega_0$ is the natural angular frequency of the system, $m$ is the mass of the membrane and $C_0=\epsilon_0\,S/d$. The membrane displacement $yd$ and membrane's velocity $\dot{y}d$ play the role of the internal state variables $x_1$ and $x_2$ (in Eqs. (\ref{Ceq1}) and (\ref{Ceq2}) $x=[x_1,x_2]$).

\subsection{Membrane memcapacitor model including membrane resistance} \label{sec22}

An experimental realization of the bistable membrane memcapacitive system based on graphene was reported in Ref.~\cite{eichler11}. It has been shown that the membrane dynamics can be described with a high precision by equations similar to Eqs. (\ref{eqC2})-(\ref{eqC3}) at relatively large oscillation amplitudes~\footnote{Ref.~\cite{eichler11} also reports a non-linear correction to the damping constant $\Gamma$ proportional to $\sim y^2$ that appears at large oscillation amplitudes and is not really important for our consideration.}. At room temperatures, however, the membrane resistance may become important (and it can also be modulated by doping the graphene sheets)~\cite{graphene2,graphene3,graphene4}. Here we consider a modified model from Sec.~\ref{sec21} taking into account a finite membrane resistance $R$ that is included in series with the membrane memcapacitance.

The external voltage $V(t)$ applied to the memcapacitive system is $V(t) = V_C(t)+V_R(t) = V_C(t)+RI(t)$, where $I(t)=\textnormal{d}q(t)/\textnormal{d}t$ is the current flowing through the memcapacitor. Eq.~\eqref{eqC1} must then be replaced by
\begin{equation}
q(t)= C(y)(V(t)-RI(t)). \tag{\ref{eqC1}a} \label{eqC1a}
\end{equation}
We define $\beta_0=2\pi / \left( \omega_0\,d \right) \sqrt{C_0 / \left( 2\,m\right)}$ , thus $\beta(t)$ in the the last term of Eq.~\eqref{eqC3} must be replaced by $\beta_0(V(t)-RI(t))$ and we can replace $I(t)$ by differentiating Eq.~\eqref{eqC1}, i.e., $I(t)=\textnormal{d}(C(y)V_C(t))/\textnormal{d}t$. From these substitutions in the last term of Eq.~\eqref{eqC3} we obtain
\begin{align}
\frac{\textnormal{d}\dot{y}}{\textnormal{d}\tau}&=-4\pi^2\,y\,\left(\left(\frac{y}{y_0}\right)^2-1\right)-\left(\Gamma+\frac{\omega_0\beta_{0}^{2}RC_{0}}{\pi\left(1+y\right)  ^{4}}V(\tau)V_{C}(\tau)\right)  \frac{\textnormal{d}y}{\textnormal{d}\tau}-\left(  \frac{\beta_{0}V(\tau)}{1+y}\right)  ^{2}\times\nonumber\\
\times&\left(  1-\frac{\omega_0RC_{0}}{\pi V(\tau)\left(  1+y\right)  }\frac{\textnormal{d}V_{C}(\tau)}{\textnormal{d}\tau}\right)  -\left(  \frac{\omega_0\beta_{0}RC_{0}}{2\pi \left(  1+y\right)  ^{2}}\right)  ^{2}\left(  \frac{\textnormal{d}V_{C}(\tau)}{\textnormal{d}\tau}-\frac{V_{C}(\tau)}{(1+y)}\frac{\textnormal{d}y}{\textnormal{d}\tau}\right)  ^{2} \tag{\ref{eqC3}a} .\label{eqC3a}
\end{align}
We note that in the case of nano-scale memcapacitors (for example with applications in VLSI circuits), both $R$ and $C_0$ are very small, so we can safely neglect the terms in $(RC_0)^2$.
Therefore, the above equation clearly shows that the resistance $R$ increases the effective damping coefficient (i.e., the term multiplying $\textnormal{d}y/\textnormal{d}t$). In fact, the qualitative results from equations \eqref{eqC1a} and \eqref{eqC3a} are not different from these based on Eqs. \eqref{eqC1} and \eqref{eqC3}. Therefore, for the sake of simplicity, only the simulation results based on the model without the resistance are presented below.

\subsection{READ and WRITE operations} 

In order to perform logic functions, the circuit architecture should support single device READ/ WRITE operations as well as computing -- the collective dynamics of coupled devices. This paper focuses on the computing functionality. Single device operations were considered in previous work \cite{pershin11c}. The READ process involves a capacitance measurement, which can be performed by any suitable technique \cite{petkov04a}. In particular, in our previous work \cite{pershin11c} it was suggested to use differential voltage amplifiers (similar to those used in DRAMs) for reading purposes. The same technique could be employed to read the capacitance of the membrane memcapacitive systems. Charged to the same voltage, the smaller capacitance will result in a weaker current (below the threshold of the differential voltage amplifier) that could be distinguished from the stronger one (above the threshold).

Here, we just briefly review the more involved WRITE process. We note that in the presence of an applied voltage, the capacitor plates experience an attractive force toward each other. Correspondingly,
the double well potential from Fig.~\ref{fig1} becomes asymmetric -- its right minimum moves up while the left one moves down. At a certain voltage magnitude, the right minimum disappears and the system, regardless of its initial state, ends up in the left minimum. This is the basis to set the system to '1'. In order to set the system to '0', a higher voltage is needed. When such voltage is applied and then removed, an accumulated elastic energy becomes sufficiently strong to overcome the potential barrier and set the system to '0' (see Ref. \cite{pershin11c} for more details).

\section{Logic gate} 

\begin{figure}[tb]
 \begin{center}
  \includegraphics[width=6.7cm]{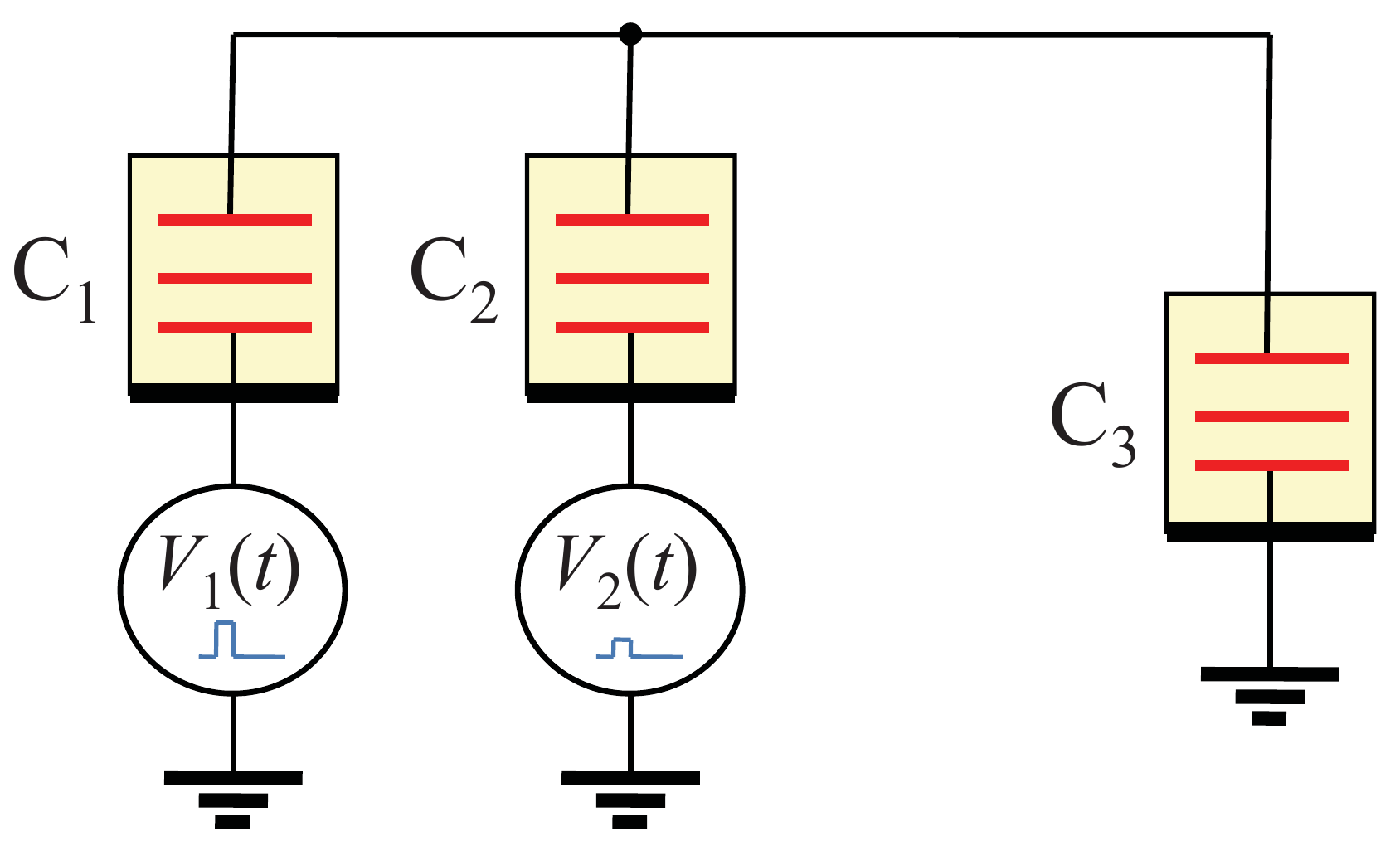}
\caption{\label{fig2} Circuit considered in this work. Here, memcapacitive systems C$_1$ and C$_2$ hold
input values, while C$_3$ the output one. Two voltage sources are used to subject the input memcapacitive systems C$_1$ and C$_2$ to pulse sequences $V_1(t)$ and $V_2(t)$.}
\end{center}
\end{figure}

We consider first logic operations with the circuit shown in Fig. \ref{fig2}. This circuit involves three memcapacitive systems and two voltage sources and can be considered as a sub-part of a larger circuit, effectively decoupled from this larger circuit with appropriate switches. Such a larger circuit could be, for example, similar to the DCRAM architecture we have previously
introduced \cite{traversa14a}. In fact, like in our previous work \cite{traversa14a} with solid-state memcapacitive systems \cite{martinez09a}, we expect more cells will provide a larger set of logic gates at different voltages.

For the sake of simplicity, we do not show any initialization and measurement setup in Fig.~\ref{fig2} since only the computing stage is of interest. All results reported in this paper were obtained utilizing completely overlapping single square pulses $V_1(t)$ and $V_2(t)$, as shown in the top panel of Fig.~\ref{fig3}.

The circuit dynamics is found using Kirchhoff's circuit laws together with Eqs. (\ref{eqC1})-(\ref{eqC3}) defining the response and dynamics of memcapacitive devices. In particular, one can find that at each moment of time the voltages across the three memcapacitive systems are given by
\begin{eqnarray}
V_{C_1}&=&\frac{C_2V_2(t)-(C_2+C_3) V_1(t)}{C_1+C_2+C_3}, \label{VC1} \\
V_{C_2}&=&\frac{C_1V_1(t)-(C_1+C_3) V_2(t)}{C_1+C_2+C_3}, \label{VC2} \\
V_{C_3}&=&\frac{C_1 V_1(t)+C_2V_2(t)}{C_1+C_2+C_3}, \label{VC3}
\end{eqnarray}
where the voltages are defined with respect to the terminal denoted by the thick line in the memcapacitive system symbol in Fig.~\ref{fig2}. These instantaneous values of voltages influence the dynamics of the internal state variables through Eq. \ref{eqC3}. In what follows, it is assumed that the input values are stored in C$_1$ and C$_2$, while C$_3$ is reserved for the output value. However, for certain regions of pulse parameters, the final states of C$_1$ and C$_2$ are different from the input ones. Therefore, these two memcapacitive systems could also be used to store the computing result in some cases and thus few different logic operations could be realized in a single shot (see also Sec. \lq\lq{}Reduced circuit\rq\rq{}).

\subsection{Material implication} 

\begin{figure}[tb]
 \begin{center}
  \includegraphics[width=0.48\textwidth]{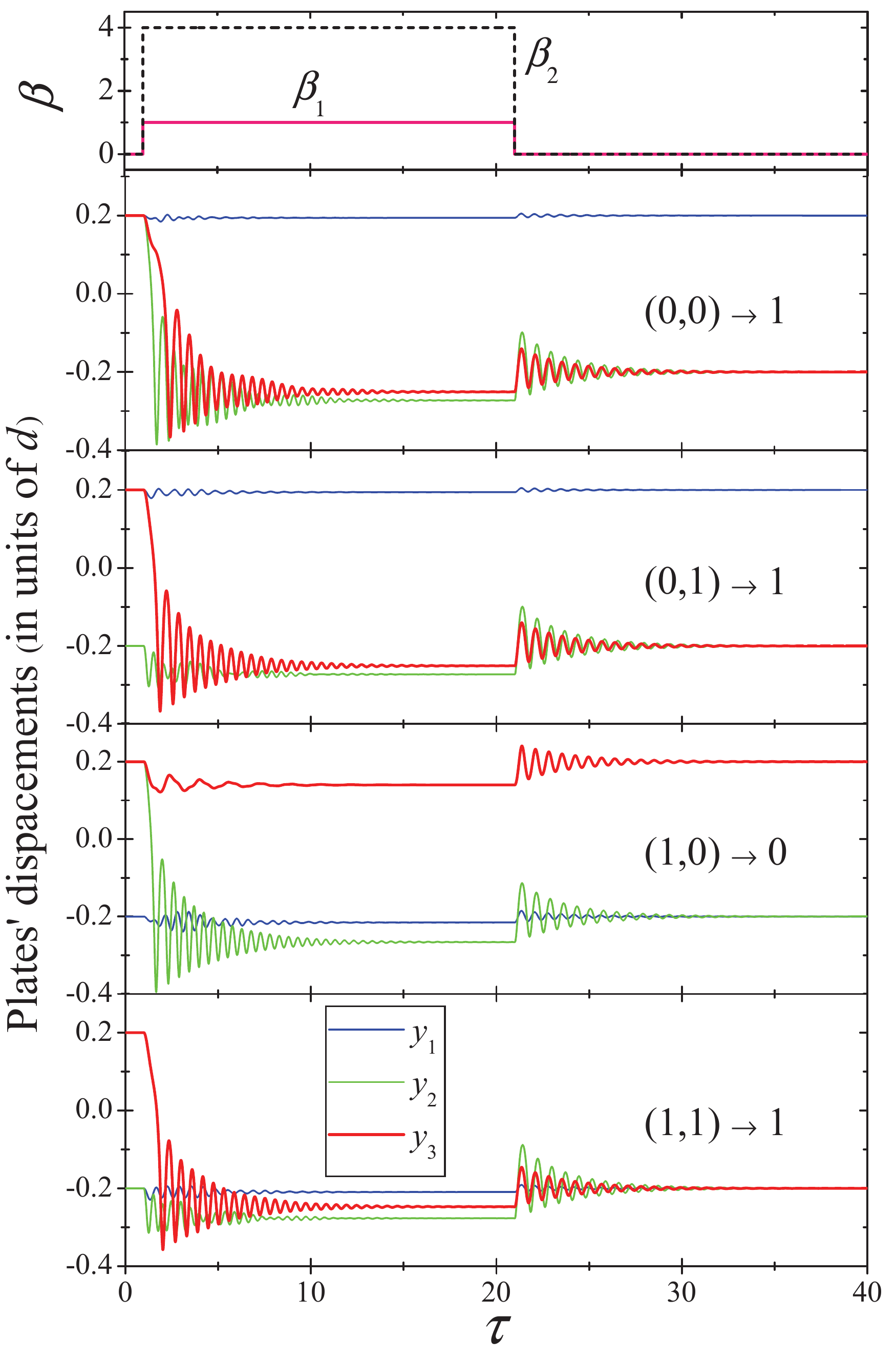}
\caption{\label{fig3} Material implication (C$_1\rightarrow$C$_2$)$=$C$_3$ with membrane memcapacitive systems. The four bottom plots show dynamics of coupled memcapacitive systems (according to Fig.~\ref{fig2} circuit configuration) at different initial conditions. The voltage pulses are demonstrated in the top plot. The initial state of C$_3$ is always '0'. These plots were obtained using the parameter values $\Gamma=0.7$ and $y_0=0.2$.}
\end{center}
\end{figure}

We are now ready to show that these membrane memcapacitive systems are able to perform logic operations. To do this we focus on the logic material implication previously demonstrated with memory resistive devices \cite{borghetti10a}. Material implication is a very important logic function because it can be used to synthesize the negation (with the help of a false operation), which, together with implication, allows for a functionally complete set of logic gates.

\begin{table}[!tb]
\renewcommand{\arraystretch}{1.3}
\caption{Codes of certain logic operations calculated according to Eq.~(\ref{code}). These codes are defined with respect to different pairs of input values (C$_1$,C$_2$).
For example, NOT C$_1$ is the logical negation on C$_1$, copy C$_2$ is the copy of the input state of C$_2$ into the final state of a given system, IMP$_1$ is the material implication C$_1\rightarrow$C$_2$, etc. More details are given in the text.}
\label{codes}
\centering
\begin{tabular}{|c|c||c|c|}
\hline
set to 0      & 0  & AND      & 8 \\
\hline
NOR        & 1  & NOT(XOR)  & 9 \\
\hline
NOT(IMP$_2$) & 2  & copy C$_2$ & 10 \\
\hline
NOT C$_1$   & 3  & IMP$_1$   & 11\\
\hline
NOT(IMP$_1$) & 4  & copy C$_1$ & 12 \\
\hline
NOT C$_2$    & 5  & IMP$_2$   & 13 \\
\hline
XOR         & 6  & OR      & 14\\
\hline
NAND        & 7  & set to 1   & 15\\
\hline
\end{tabular}
\end{table}

In order to demonstrate the material implication, let us consider the circuit dynamics at specific amplitudes of voltage pulses $\beta_1(\tau)$ and $\beta_2(\tau)$, namely, $\beta_1=1$ and $\beta_2=4$. Fig.~\ref{fig3} shows the dynamics of the internal states of memcapacitive systems (the position of the flexible plate) for four possible initial states of C$_1$ and C$_2$. It is assumed that C$_3$ is in `0' state at $\tau=0$. Clearly, C$_3$ remains in '0' only at the (1,0) input combination and its final state is `1' for all other input combinations. This is the material implication. We also note that the final state of C$_2$ is always `1'. The state of C$_1$ remains unchanged during the circuit dynamics operation.

\subsection{Map of logic operations} 

In order to better understand which logic operations can be implemented with the memcapacitive logic circuit from Fig.~\ref{fig2}, we prescribe a numerical value to operation results as follows. Taking $w_i=1,2,4,8$ as weights for the input combinations (0,0), (0,1), (1,0) and (1,1), a numerical code is calculated as a weighted sum of the final state of a particular memcapacitive system, namely,
\begin{equation}
\textnormal{code}=\sum\limits_{i=1}^4 w_i b_{ij}^f,
\label{code}
\end{equation}
where $b_{ij}^f$ is the final state (0 or 1) of the device of interest $j$ (C$_1$, C$_2$ or C$_3$) for $i$-th input combination (0,0), (0,1), (1,0) or (1,1) that correspond to $i=1,2,3,4$. For example, for the material implication function shown in Fig.~\ref{fig3}, the code for the final state of C$_3$ is $1\cdot 1+2\cdot 1+4\cdot 0+8\cdot 1=11$. Therefore, 11 is the code for material implication C$_1\rightarrow$C$_2$. Similarly, one can find that the material implication C$_2\rightarrow$C$_1$ corresponds to the code value 13. Table~\ref{codes} summarizes the codes for all operations implemented with membrane memcapacitive logic.

\begin{figure}[tb]
 \begin{center}
  \includegraphics[width=0.48\textwidth]{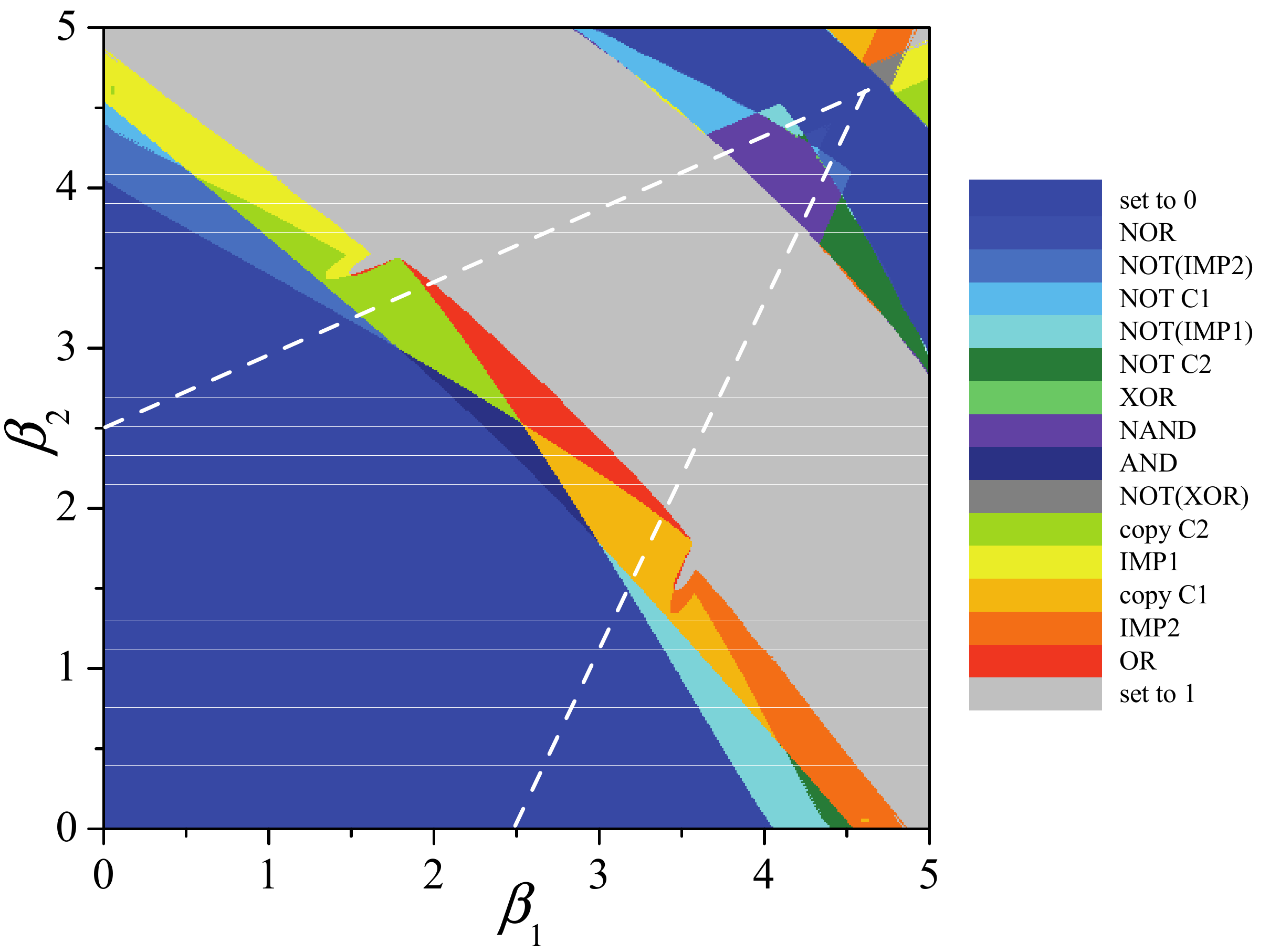}
\caption{\label{fig4} Logic operation type as a function of pulse amplitudes $\beta_1$ and $\beta_2$ for the output memcapacitive system C$_3$. Each point of this plot was obtained with a calculation similar to that shown in Fig.~\ref{fig3}. This plot was obtained using the pulse width $T=20$, $\Gamma=0.7$, $y_0=0.2$. The region between two white dashed lines corresponds approximately to the operation regime such that C$_1$ and C$_2$ stay unchanged.}
\end{center}
\end{figure}

Fig.~\ref{fig4} is the main result of this work. It identifies the regions of voltage pulse amplitudes $\beta_1$ and $\beta_2$ realizing specific logic functions as
the final state of C$_3$. Each point of this plot is calculated similarly to Sec. \lq\lq{}Material implication\rq\rq{} calculation assuming that C$_3$ is in 0 at $t=0$. As expected,
at smaller values of $\beta_1$ or $\beta_2$ and any input combination, the final state of C$_3$ is 0. Material implication, OR, NAND and some other functions are found at higher
values of the applied pulses as shown in Fig.~\ref{fig4}. This calculation demonstrates that the same memcapacitive circuit is capable of realizing different logic functions on demand without any changes in the circuit configuration. The circuit thus performs {\it polymorphic} computing in the sense discussed in Ref. \cite{traversa14a}.

\begin{figure}[tb]
 \begin{center}
  \includegraphics[width=0.48\textwidth]{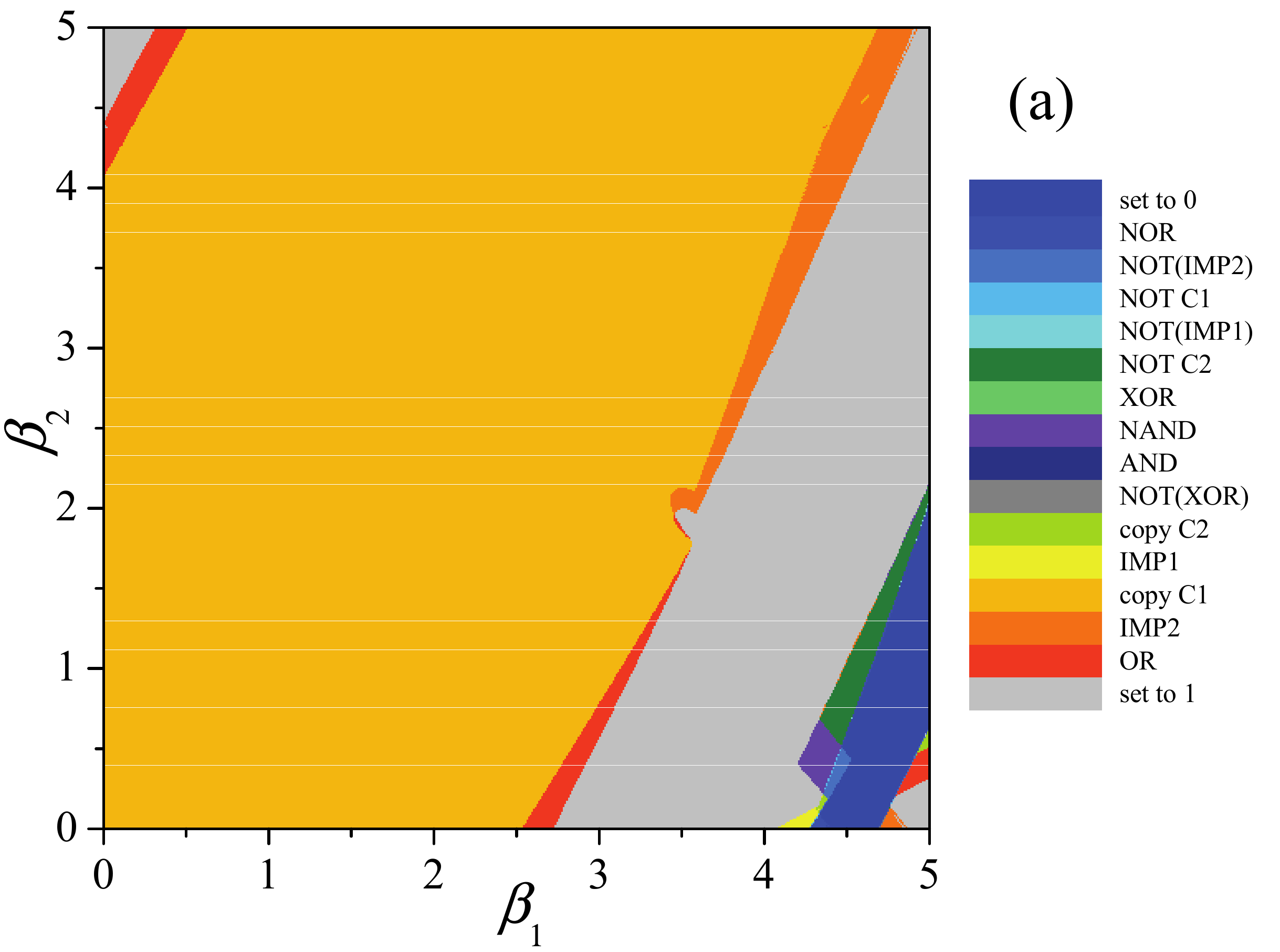}
  \includegraphics[width=0.48\textwidth]{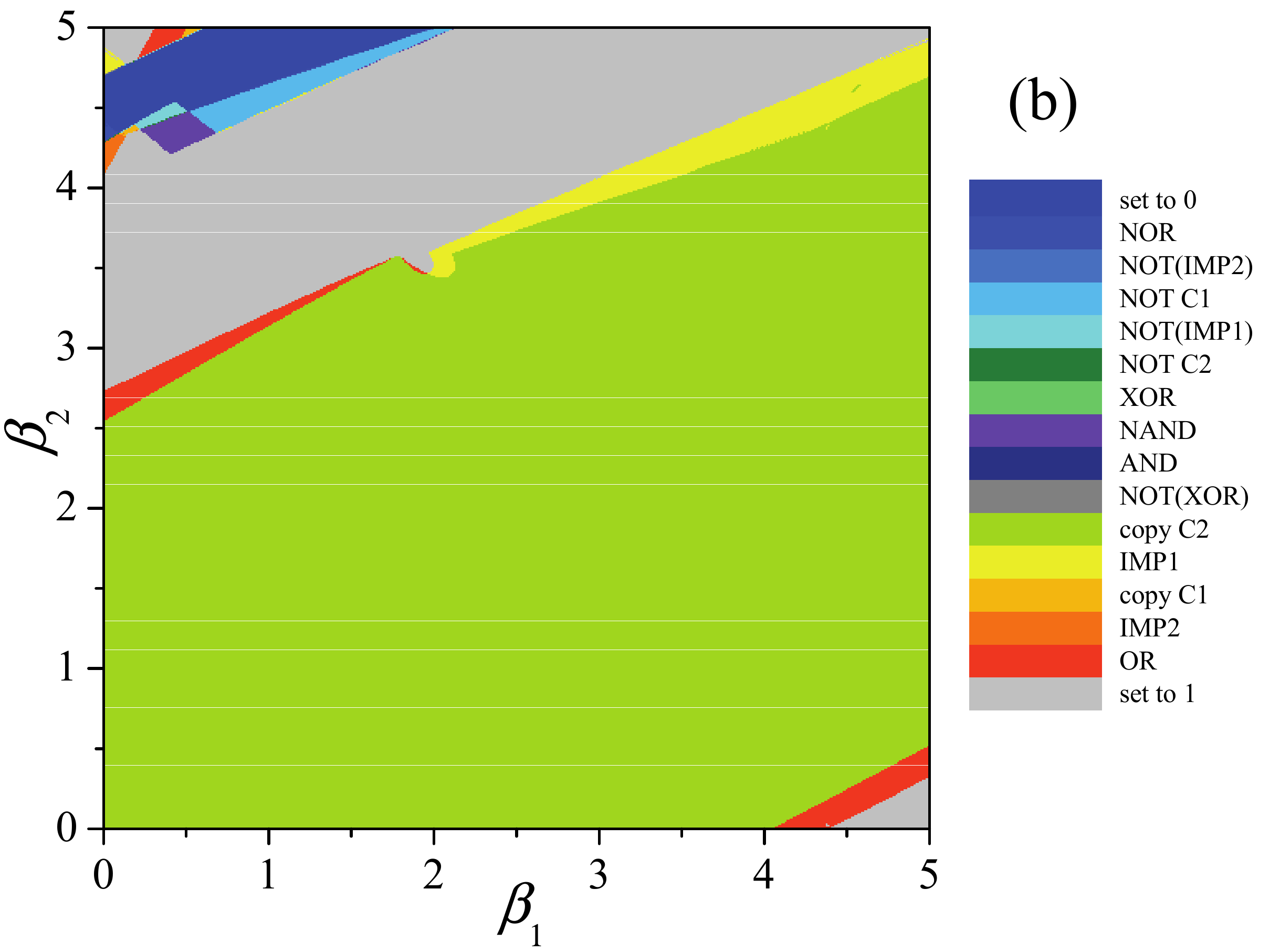}
\caption{\label{fig5} Logic operation type as a function of pulse amplitudes $\beta_1$ and $\beta_2$ for the input memcapacitive systems C$_1$, (a), and C$_2$, (b). The data for these plots and Fig.~\ref{fig4} were obtained within the same calculations.}
\end{center}
\end{figure}

The final states of C$_1$ and C$_2$ are shown in Fig.~\ref{fig5}. Although there are large regions where C$_1$ and C$_2$ stay unchanged, one can identify regions of amplitudes implementing the material implication (codes 11 and 13), `set to 1' (code 15) and some other functions. Therefore, a logic function and initialization or two different logic functions can be performed in a single step (intrinsic parallelism \cite{diventra13a,traversa14b}) thus further increasing the efficiency of calculations.

\begin{figure}[tb]
 \begin{center}
  \includegraphics[width=0.48\textwidth]{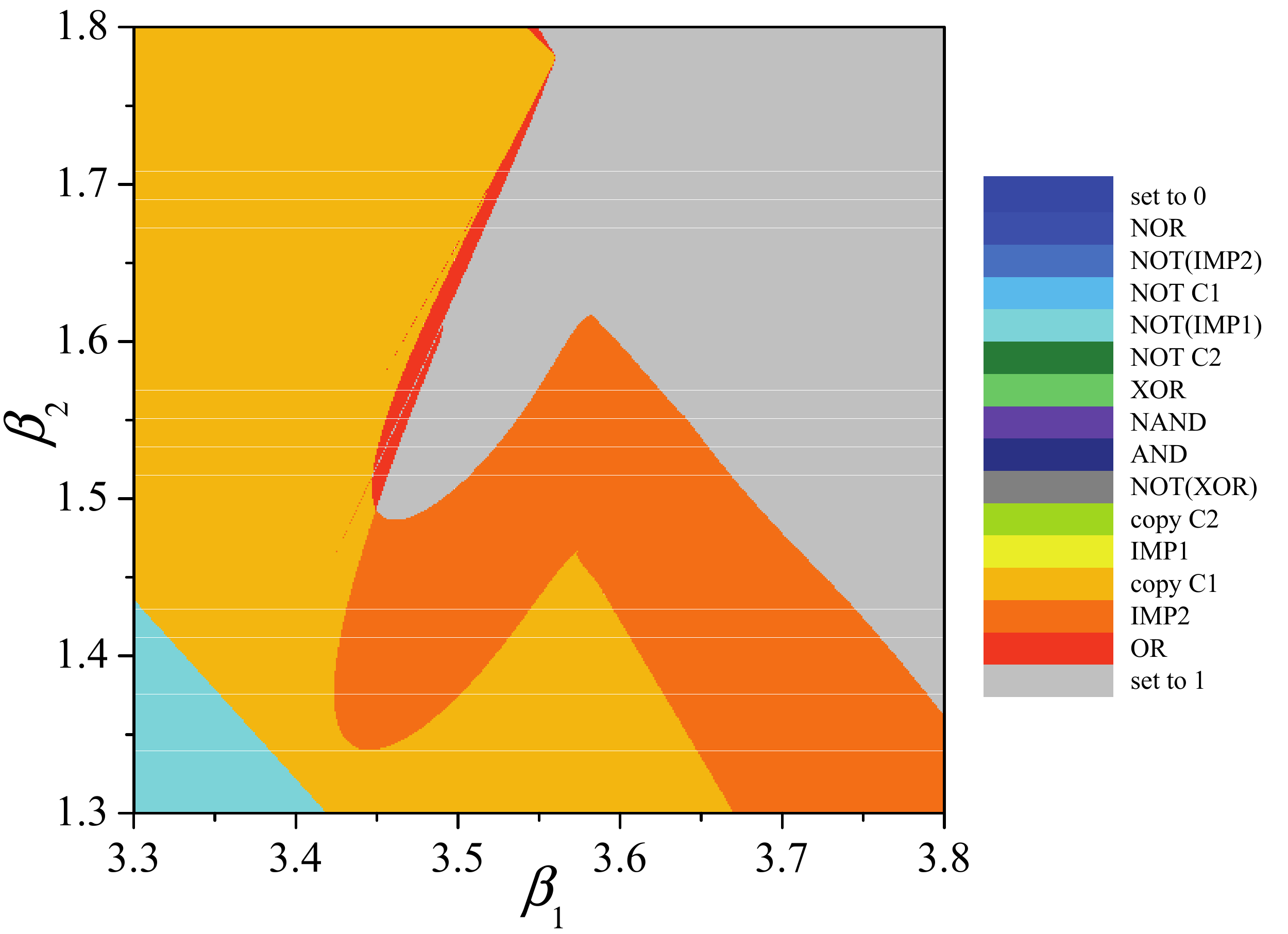}
\caption{\label{fig56} Magnified region of Fig.~\ref{fig4} showing non-trivial fine structures.}
\end{center}
\end{figure}

We also note that there are regions in the input-voltage parameter phase space that show non-trivial features such as those two irregularities observed at approximately ($\beta_1=1.5$, $\beta_2=3.5$) and ($\beta_1=3.5$, $\beta_2=1.5$) in Fig.~\ref{fig4}. Figure~\ref{fig56} shows the feature magnification revealing additional fine structures (straight lines) in those regions. These features originate from the complex dynamics of memcapacitive systems already spotted in a previous work of one of us \cite{pershin11c}.
We emphasize that such irregularities are observed only in limited intervals of voltage amplitudes, and therefore can be easily avoided in practical realizations of membrane memcapacitive systems. Moreover, by varying geometrical ($y_0$) and physical ($\Gamma$) parameters as shown in the Supporting Information, the chaotic behavior of the circuit can be minimized.

\subsection{Reduced circuit} 

\begin{figure}[tb]
 \begin{center}
  \includegraphics[width=0.27\textwidth]{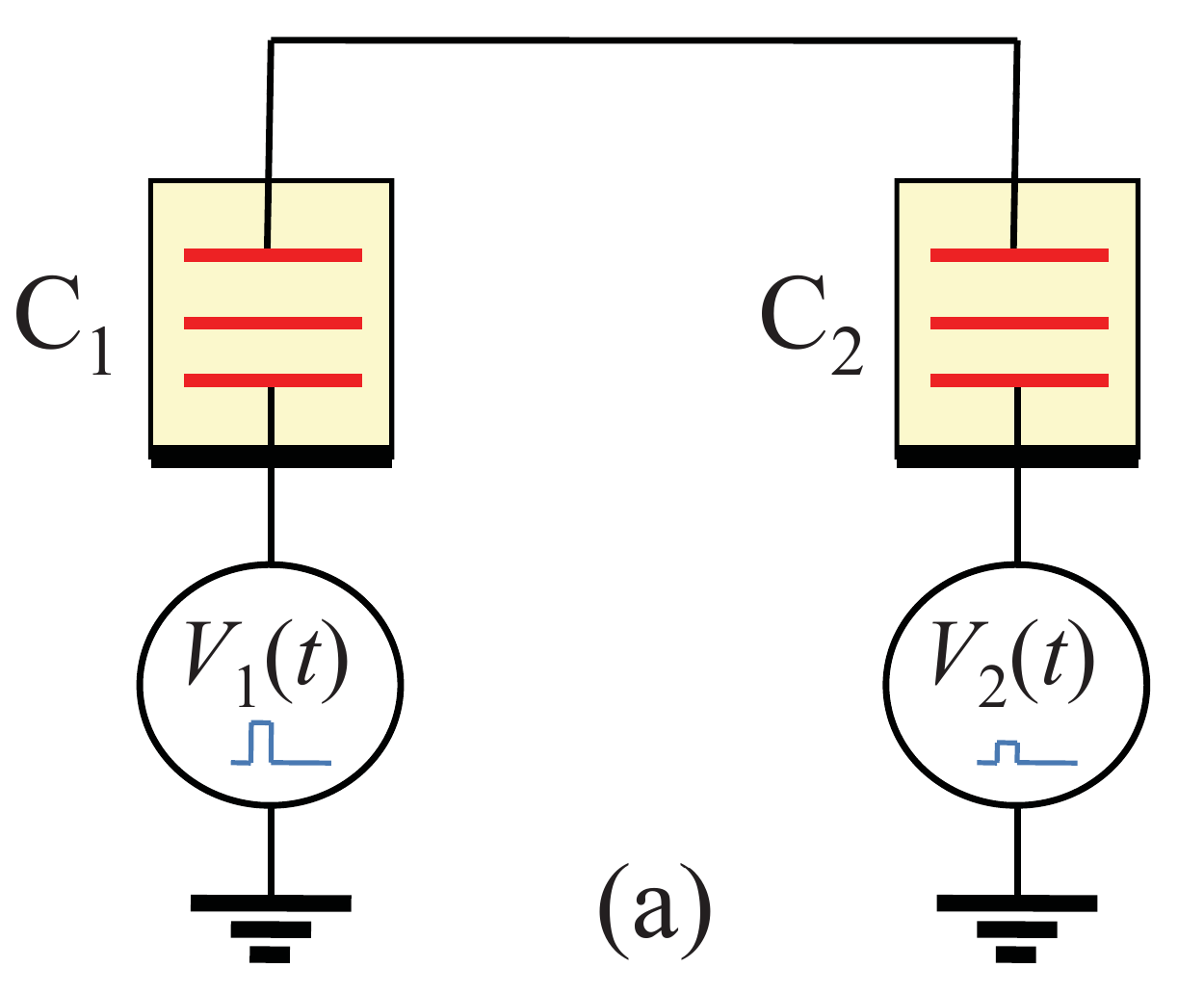}
  \includegraphics[width=0.48\textwidth]{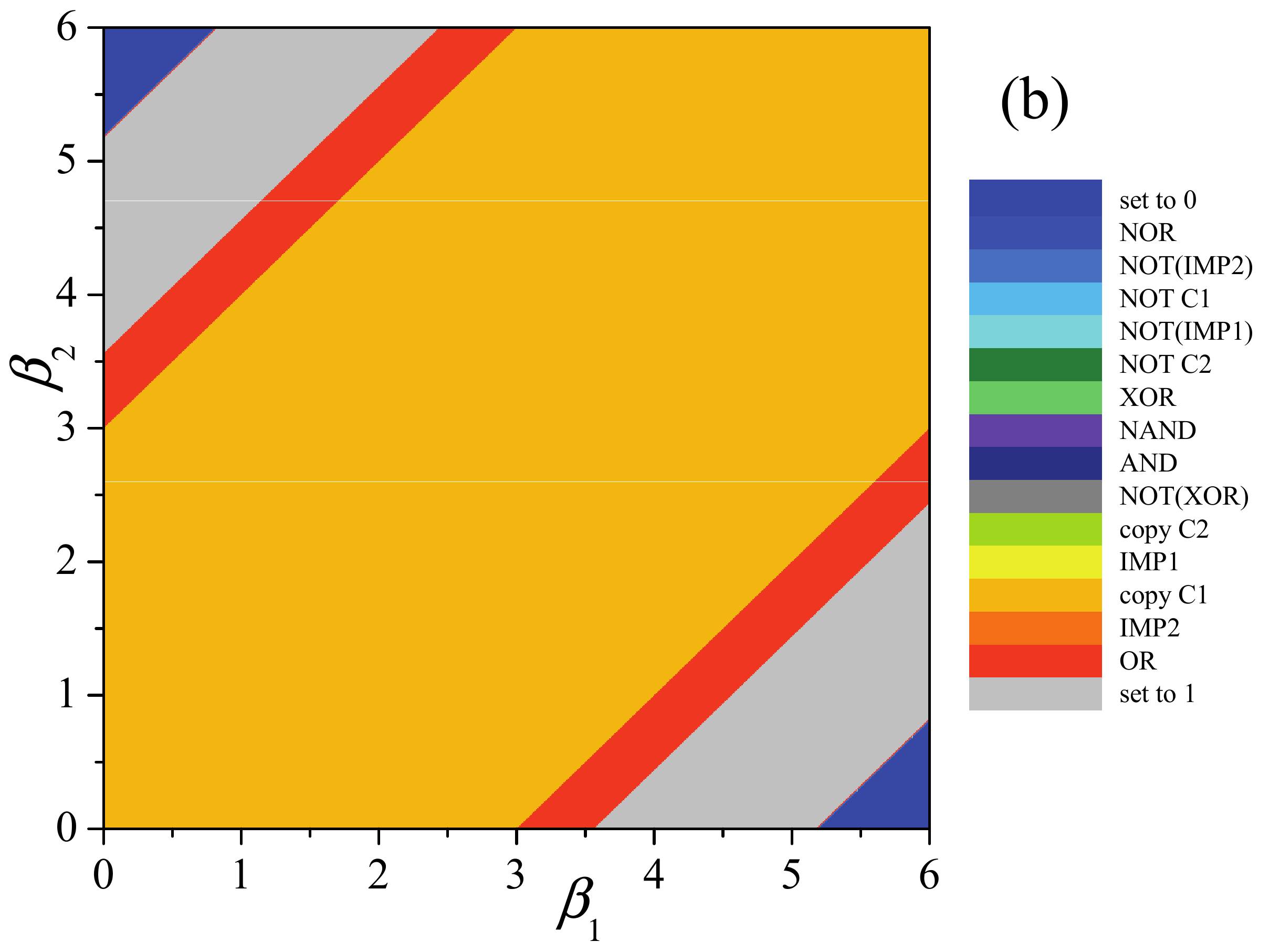}\\
   \includegraphics[width=0.45\textwidth]{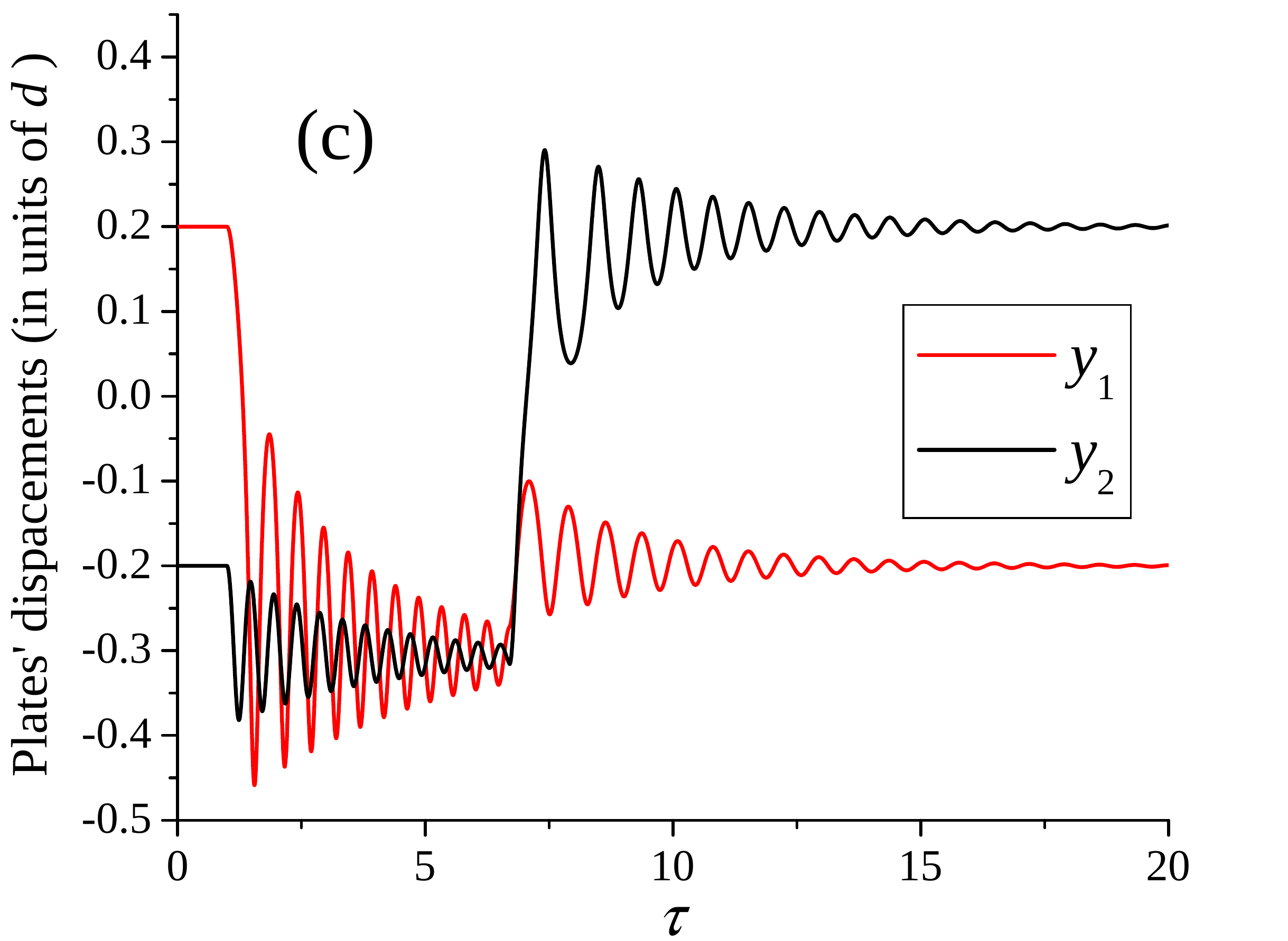}
  \includegraphics[width=0.45\textwidth]{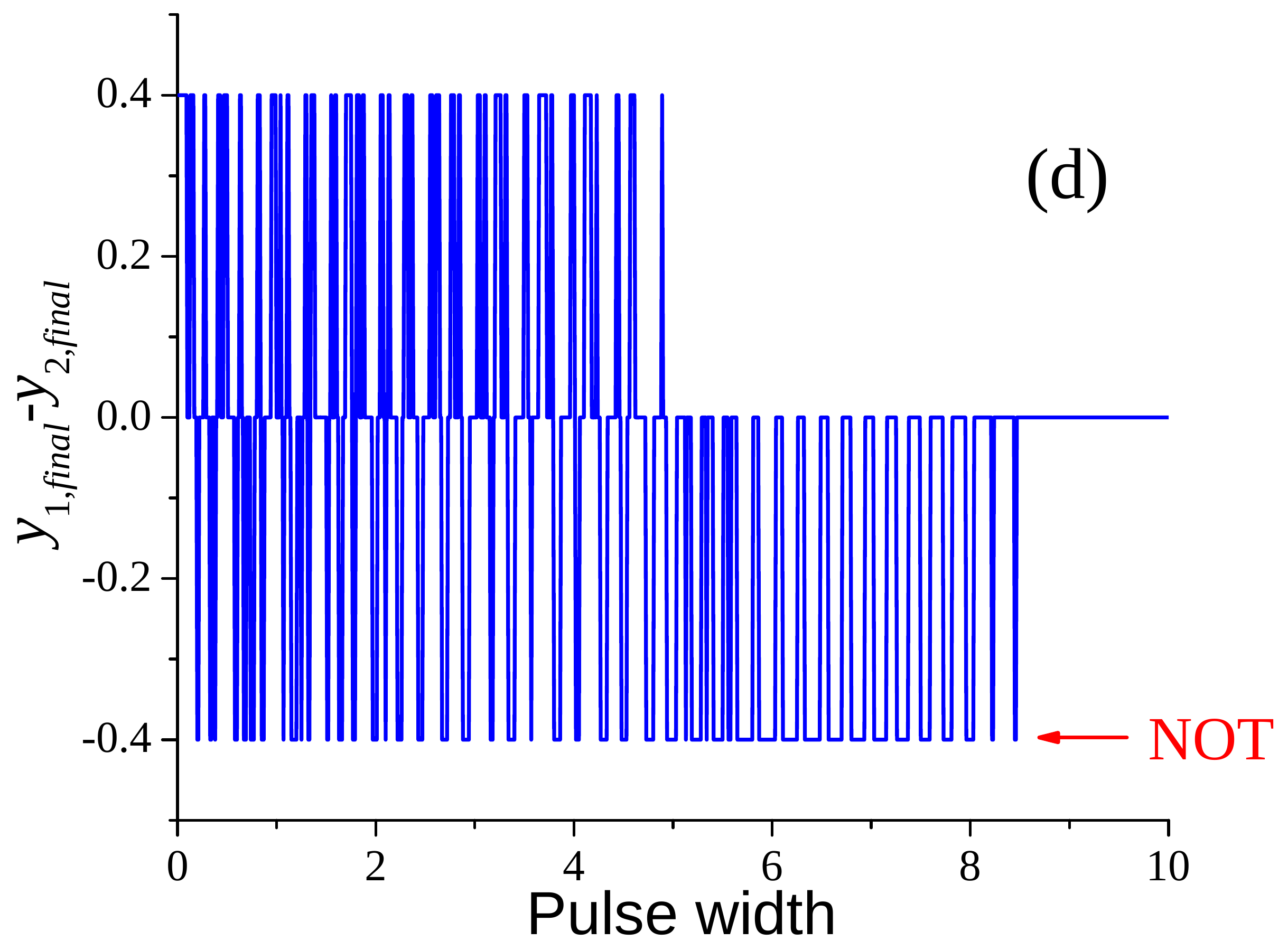}
\caption{\label{fig6} (a) Reduced circuit layout. (b) Logic operation type as a function of pulse amplitudes $\beta_1$ and $\beta_2$ calculated for the final state of C$_1$. (c) Demonstration of the NOT gate with a single memcapacitive system: $0\rightarrow 1$ and $1\rightarrow 0$ when the system is subjected to the same pulse of $\beta=2.8$ magnitude and $T=5.7$ duration. (d) Final $y_1-y_2$ as a function of the pulse width at the same value of $\beta=2.8$ showing intervals of the NOT gate.}
\end{center}
\end{figure}

The results presented in Fig.~\ref{fig5} demonstrate that the same memcapacitive device could store both input and output logic values. In order to better understand this capability we consider a reduced circuit consisting of two memcapacitive systems as sketched in Fig.~\ref{fig6}(a). We have performed simulations of the circuit dynamics subjected to the same couple of pulses and simulation parameters as we have discussed above.

The results of these simulations presented in Fig.~\ref{fig6}(b) demonstrate that even such a simple circuit is capable of implementing the OR gate in a significant interval of parameters. However, a single OR is not enough for universal computing. One possibility to attain this goal would be a combination of the OR gate and NOT gates. Considering dynamics of a single membrane memcapacitive system subjected to a voltage pulse, we have indeed found pulse parameters realizing the NOT. Fig. \ref{fig6}(c) shows an example of such realization.

In order to better understand the NOT implementation, we plot the difference of final positions of plates for different initial conditions ($y_1(0)=0.2$ and $y_2(0)=-0.2$), namely, $y_1-y_2$ at $\tau=40$. The NOT is realized when $y_1(40)=-0.2$ and $y_2(40)=0.2$. In other words, when the final $y_1-y_2=-0.4$. Fig. \ref{fig6}(d) shows multiple regions of the NOT gate, which could be achieved, for the set of system parameters selected, using a fine pulse width tuning. We emphasize that a further improvement of membrane memcapacitive logic is possible. For example, a larger set of logic operations with two memcapacitive devices could possibly be obtained adding a capacitor to the circuit in Fig.~\ref{fig6}(a). A pulse engineering is an additional opportunity that could lead to improved functionality.

\section{Impact of Device Parameters}

\begin{figure}[h]
 \begin{center}
  \includegraphics[width=0.45\textwidth]{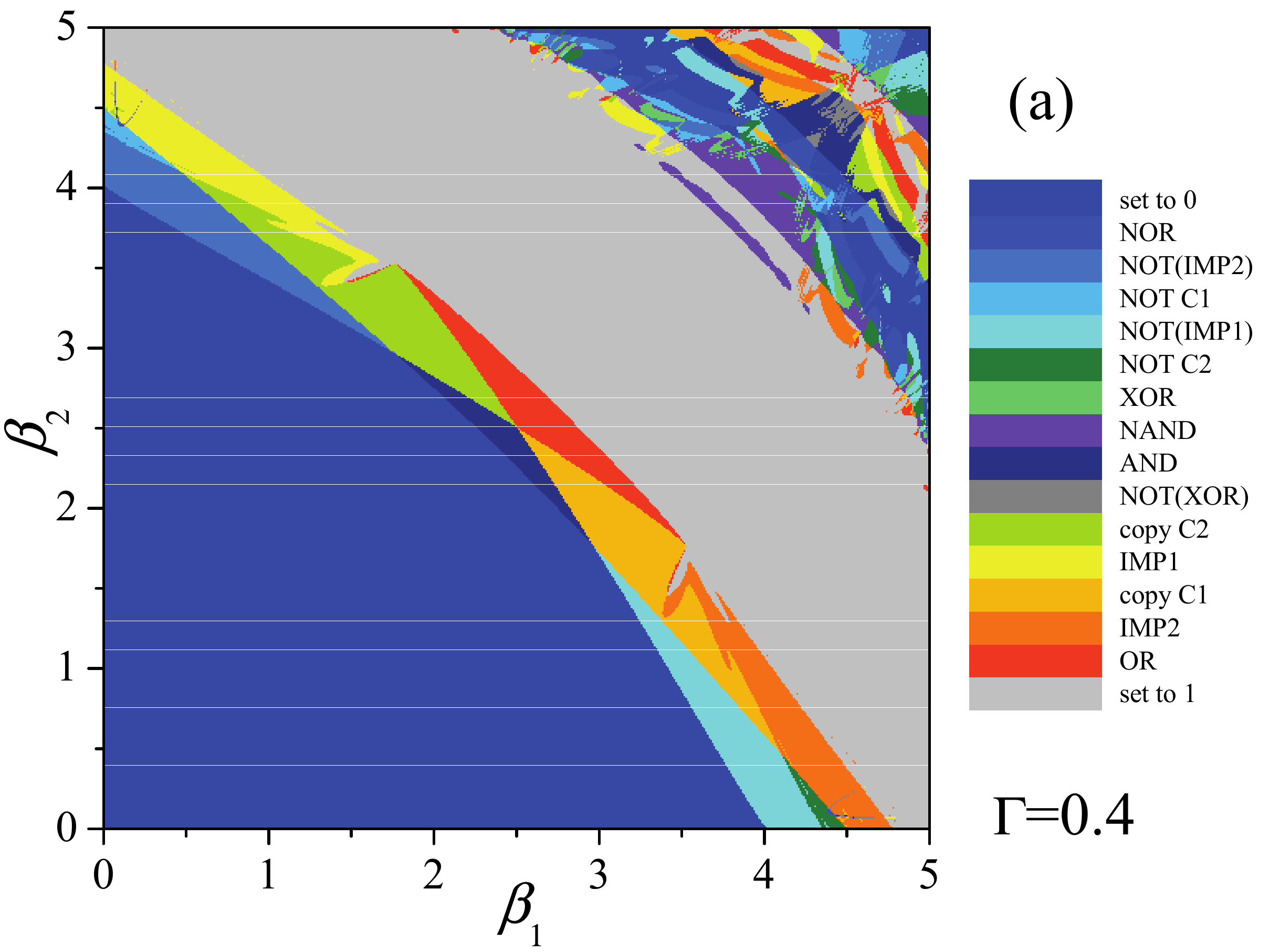}
  \includegraphics[width=0.45\textwidth]{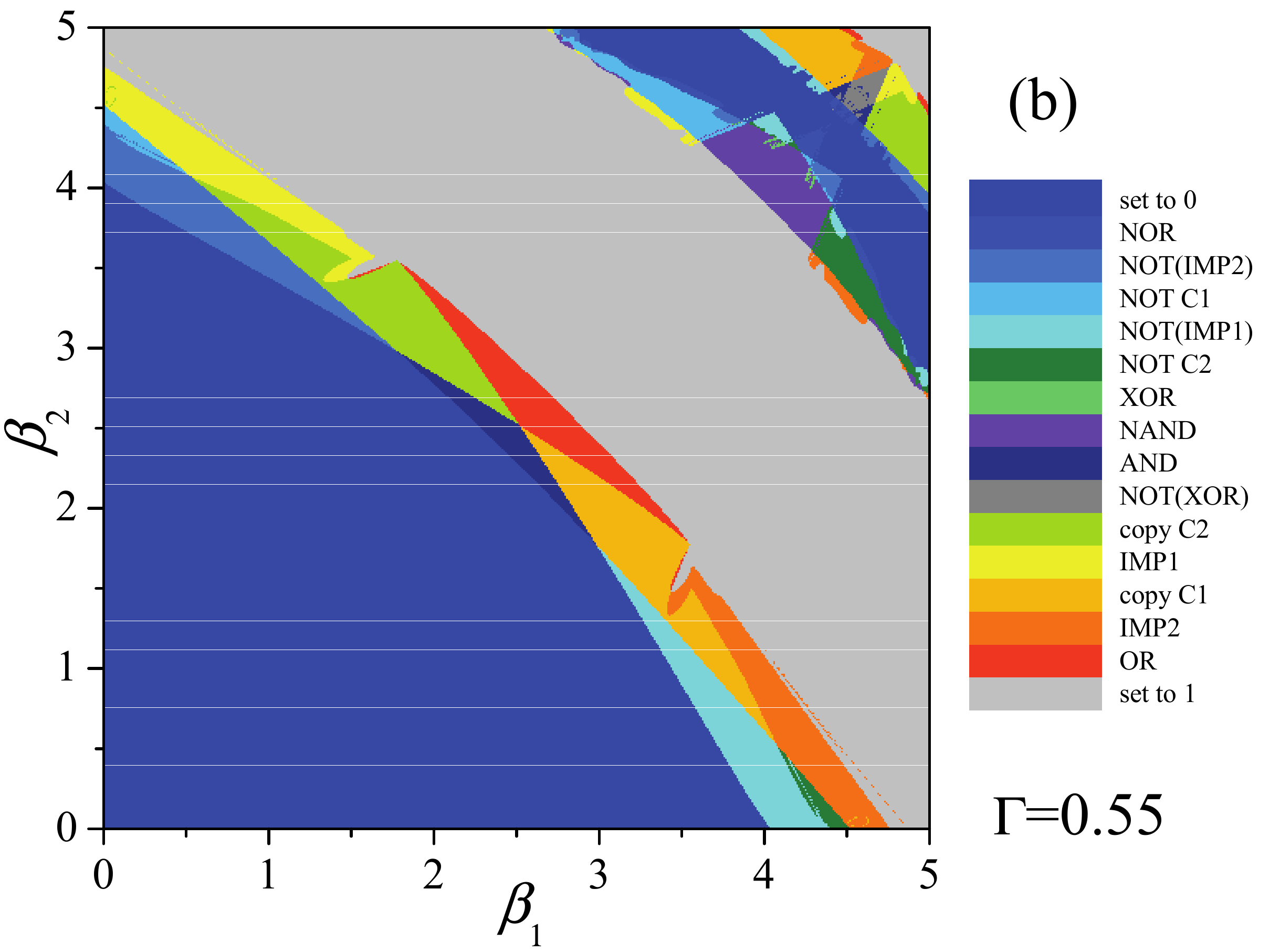}
  \includegraphics[width=0.45\textwidth]{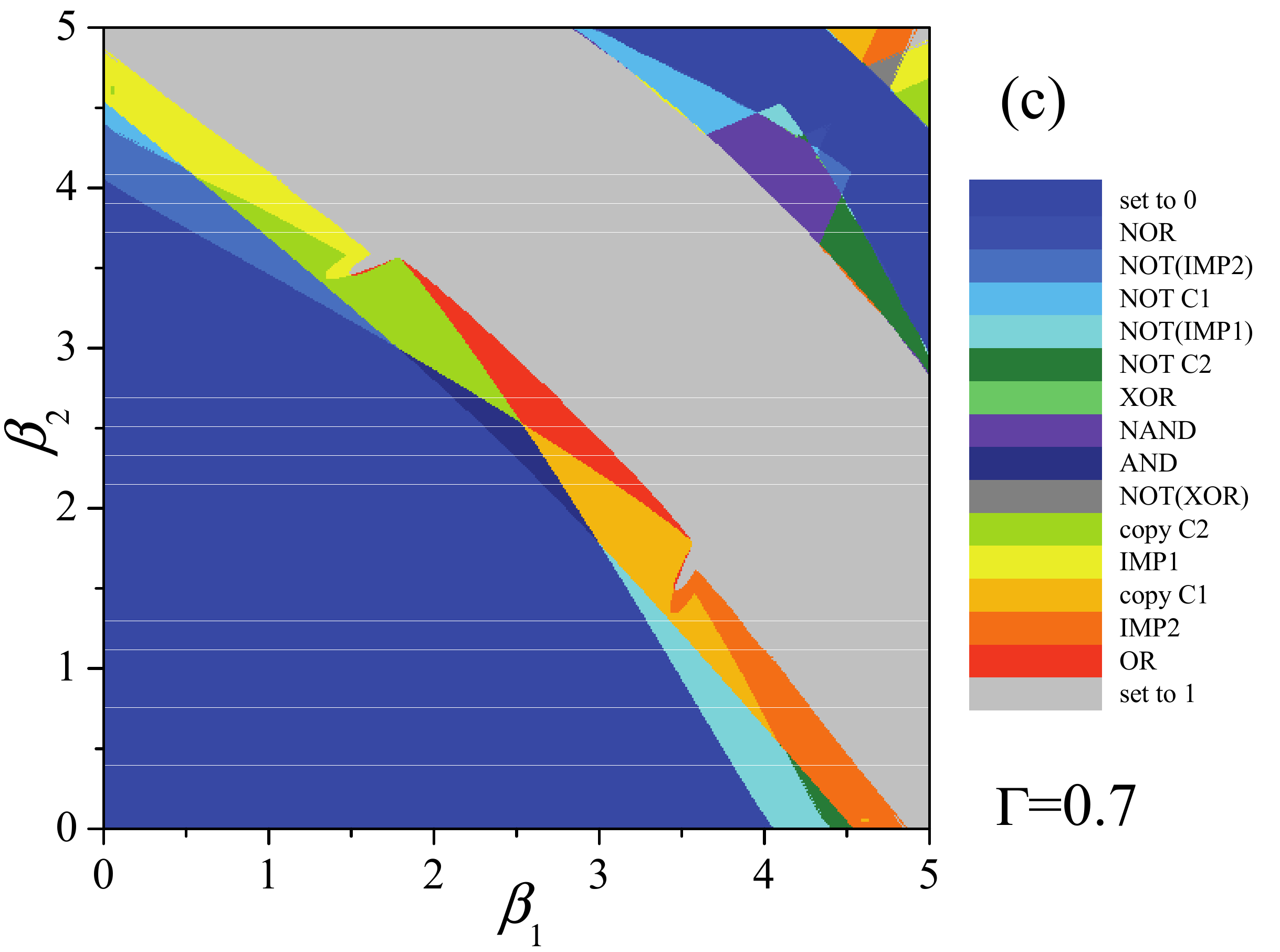}
  \includegraphics[width=0.45\textwidth]{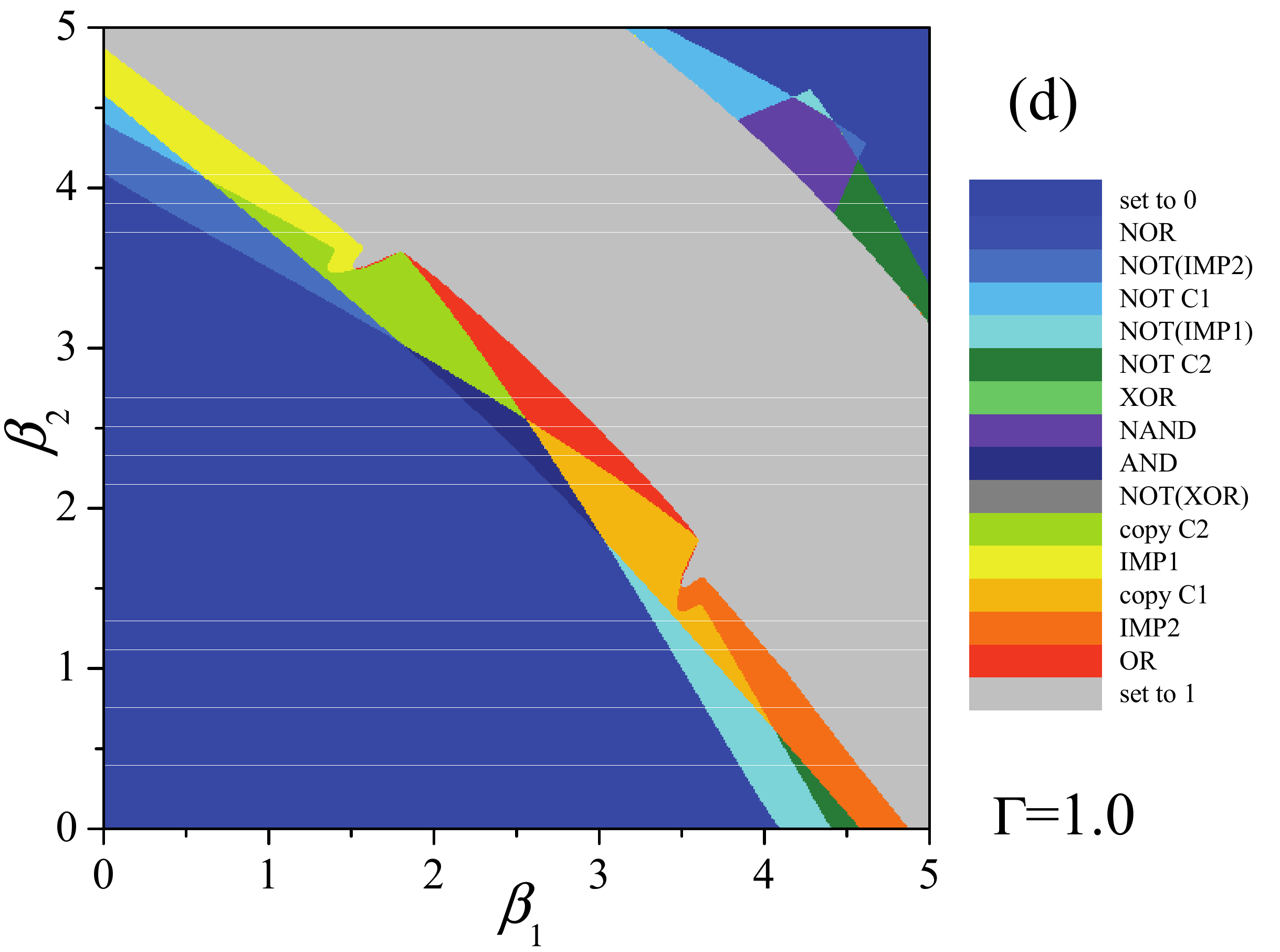}
\caption{\label{figA1} Logic operation type as a function of pulse amplitudes $\beta_1$ and $\beta_2$ for the output memcapacitive system C$_3$ for $y_0=0.2$ and several values of parameter $\Gamma$ as indicated. The calculations were performed for the circuit configuration shown in Fig. \ref{fig2} assuming two completely overlapping applied pulses of $T=20$ width.}
\end{center}
\end{figure}
In order to better understand the implementation of logic operations with membrane memcapacitive systems, we have performed several additional calculations varying parameters $y_0$ and $\Gamma$ of the model. Figure \ref{figA1} shows results of these calculations for a circuit of three memcapacitive systems (depicted in Fig. \ref{fig2}) at a fixed value of $y_0=0.2$ and several representative values of $\Gamma$. This figure demonstrates that at smaller values of $\Gamma$ (Fig. \ref{figA1}(a) and (b)) there is a significant region of chaotic-like behavior at larger values of $\beta_1$ and $\beta_2$ (see the top right parts of these plots). Increasing $\Gamma$ stabilizes this region (Fig. \ref{figA1}(c) and (d)). Clearly, already at $\Gamma=0.7$ there are no hints of uncertainty in that large $\beta_1$ and $\beta_2$ region. Therefore, while the chaotic-like behavior is already not possible at $\Gamma=0.7$, a larger value of $\Gamma$ could be used in experimental realizations of the circuit to guarantee its operation stability.

\begin{figure}[t]
 \begin{center}
  \includegraphics[width=0.45\textwidth]{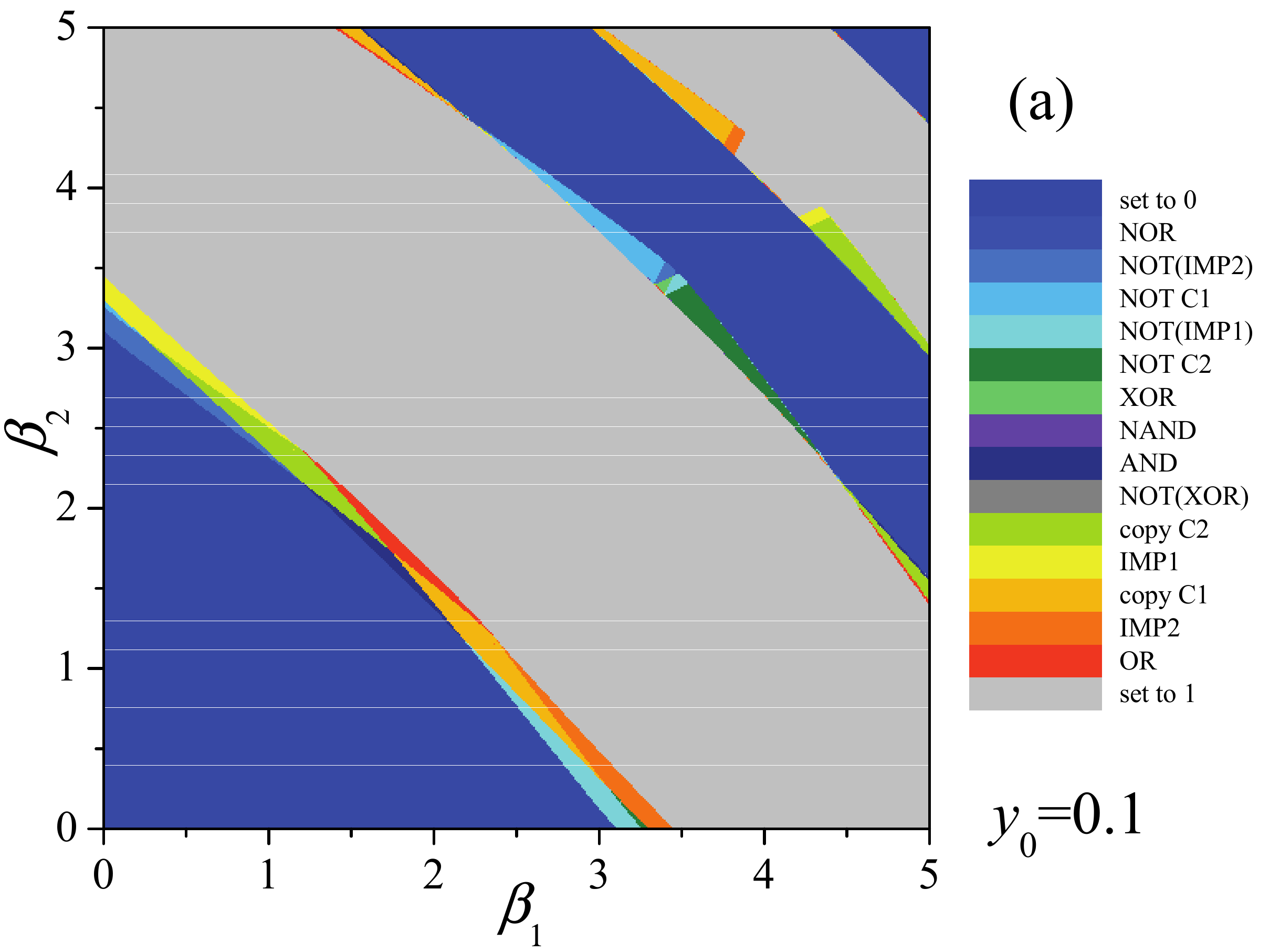}
  \includegraphics[width=0.45\textwidth]{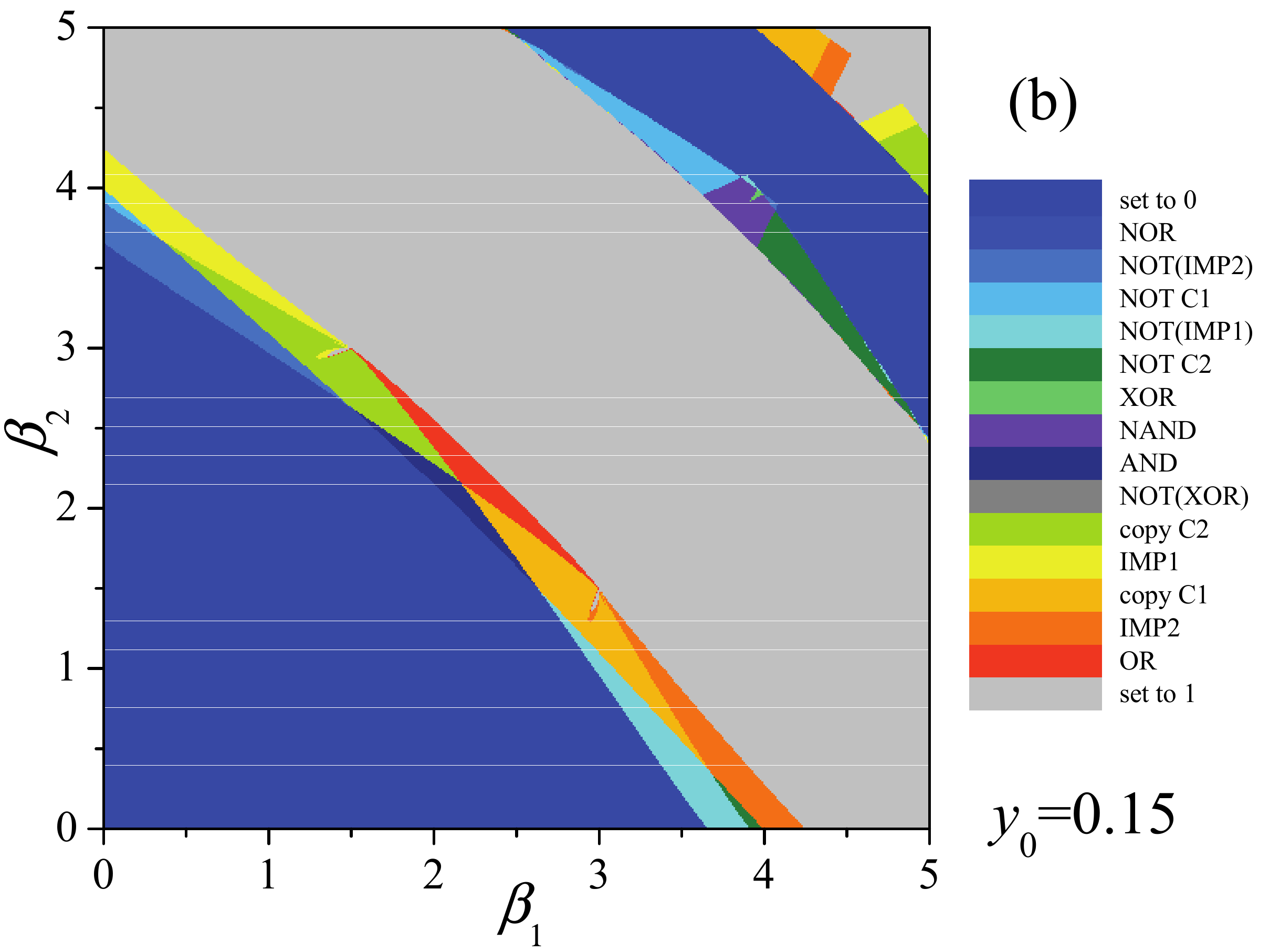}
  \includegraphics[width=0.45\textwidth]{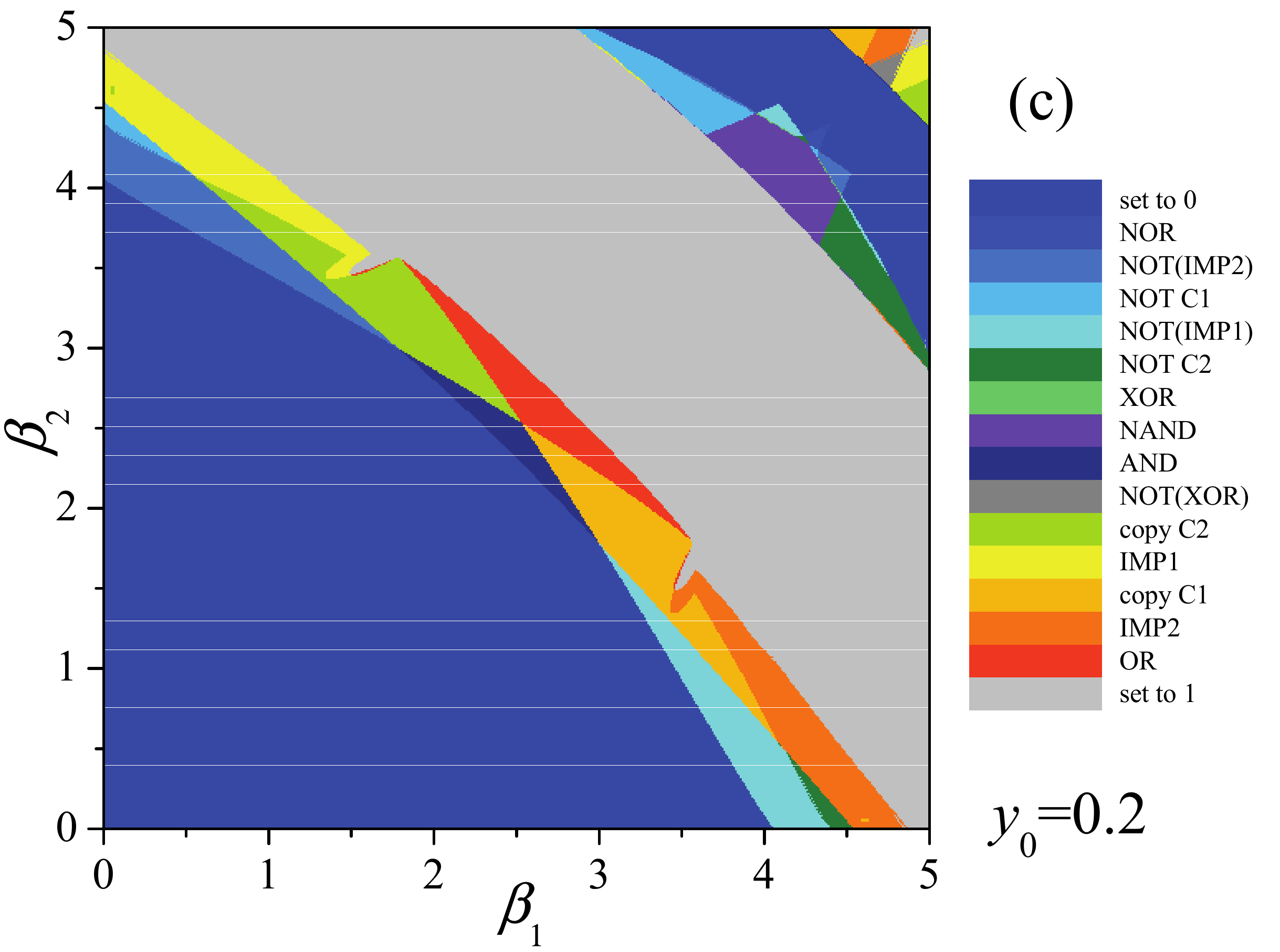}
  \includegraphics[width=0.45\textwidth]{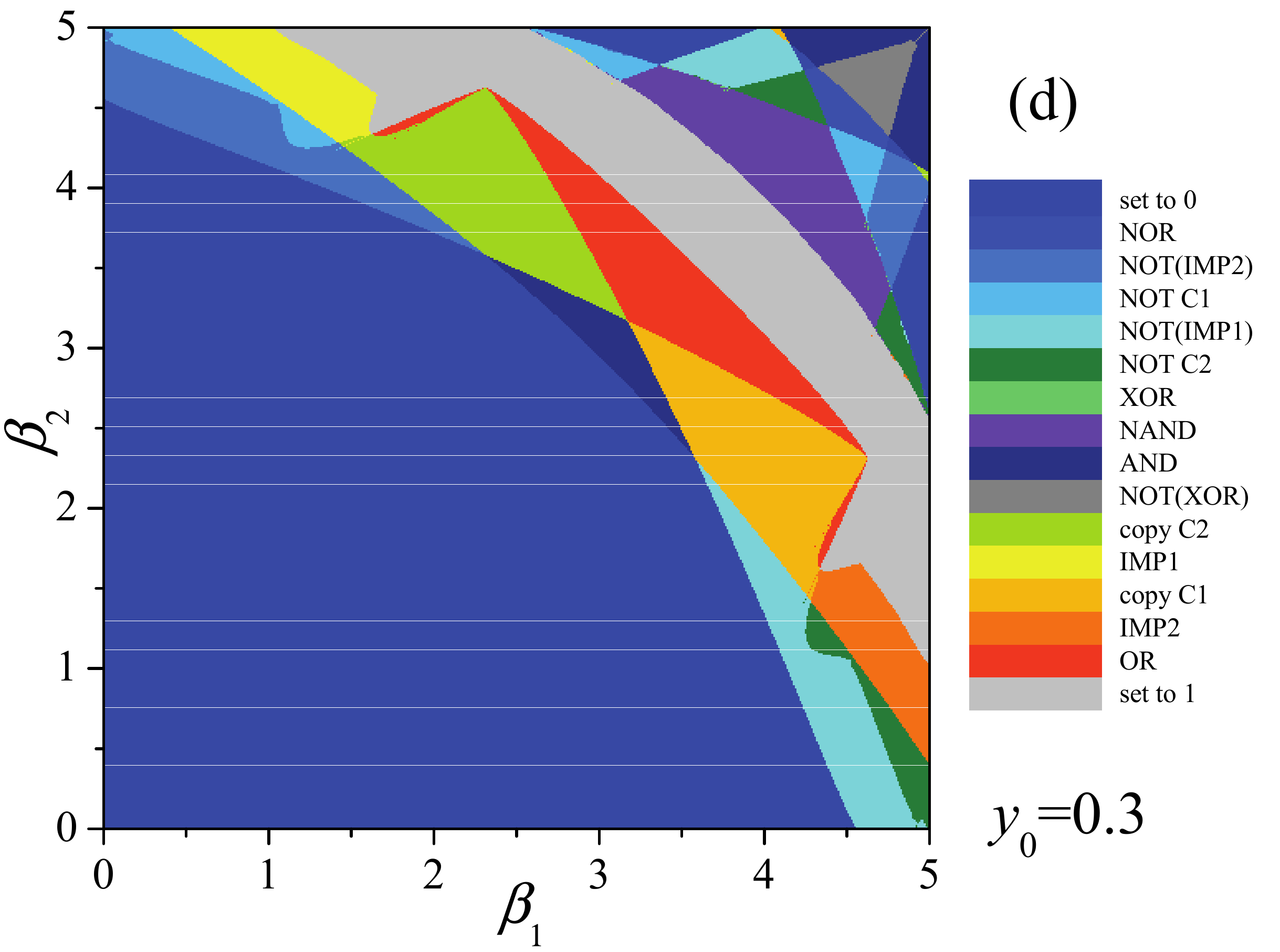}
\caption{\label{figA2} Logic operation type as a function of pulse amplitudes $\beta_1$ and $\beta_2$ for the output memcapacitive system C$_3$ for $\Gamma_0=0.2$ and several values of parameter $y_0$ as indicated. The calculations were performed for the circuit configuration shown in Fig. \ref{fig2} assuming two completely overlapping applied pulses of $T=20$ width.}
\end{center}
\end{figure}

Figure \ref{figA2} shows the results of simulations for the same three-device circuit (Fig. \ref{fig2}) at a fixed value of $\Gamma=0.7$ and several values of $y_0$. It follows from
Fig. \ref{figA2} that the regions of useful logic functions are significantly increased with increasing $y_0$. There are three potentially interesting regions in Fig. \ref{figA2}(d) (codes 7, 11, 13)
corresponding to NAND and material implications. Each of these regions provides a universal logic capability.

\section{Conclusions} 

\renewcommand{\thefootnote}{\alph{footnote}}

We have shown that memcomputing--computing {\it with} and {\it in} memory-- \cite{diventra13a} can be implemented with membrane memcapacitive systems. This demonstrates that this quite different type of memcapacitive system (compared to the solid-state memcapacitive systems previously studied \cite{traversa14a}) is also suitable for massively parallel and polymorphic computing operations directly in memory, thus offering a different realization of the memcomputing concept. Experimentally, our predictions could be verified with membrane memcapacitive systems
employing, e.g., a stressed graphene membrane\footnote{Graphene is currently used in experimental capacitors \cite{eichler11,stoller08,ElKadi13} albeit in a different role.},  a synthetic (artificial) membrane, or a molecular system as the flexible plate. Graphene membranes, for instance, have been recently demonstrated experimentally as systems for quantum information \cite{Singh14a}. Here, instead we suggest their use as semi-classical two-level systems. We thus hope our predictions will motivate further experimental and theoretical work in this direction.

From the fabrication point of view, the membrane memcapacitive memory could be realised in a CMOL-like architecture \cite{strukov05a}, which is a hybrid architecture combining a semiconductor-transistor (CMOS) layer with a layer of molecular-scale nanodevices formed between two levels of parallel nanowires \cite{strukov05a}. In our case, the top layer will be a layer of membrane memcapacitive systems coupled with the bottom CMOS layer using a set of vertical connections. It is anticipated that a single graphene-based memcapacitive system could be scaled down to few $100$s nm$^2$ area. The amount of memcapacitive systems scales almost linearly with the chip area and the most natural architecture to include circuitry to control computing, reading and writing processes is the well scalable DRAM-like architecture \cite{traversa14a} . Furthermore, membrane memcapacitive systems are passive systems (excluding control circuitry) thus the energy per operation can be as small as the energy required for the voltage pulses used to read/write and compute, that is typically of the order of few fJ for standard technologies \cite{traversa14a}. Finally, the maximum operation frequency (being passive) is directly related to the membrane damping constant $\gamma$, thus strongly depending on the materials and technologies used to build the memcapacitors, for example, how the membrane is attached to the substrate strongly affects $\gamma$. For example, from Ref.~\cite{eichler11} and from some rough estimation, a working frequency could be of the order of a hundred MHz. However, in this case very large sheets of graphene have been used and the temperature is very low. We think that at higher temperatures (where the sheet resistance is higher, see section \ref{sec22} for the impact of the resistance on the effective damping parameter) and much smaller membrane sizes, a working frequency of at least few GHz can be reached.

\section{Acknowledgment}

This work has been partially supported by the NSF grant ECCS-1202383, and the Center for Magnetic Recording Research at UCSD.




\section*{References}

\bibliography{maze}

\begin{thebibliography}{10}

\bibitem{diventra13a}
Massimiliano {Di Ventra} and Yuriy~V. Pershin.
\newblock The parallel approach.
\newblock {\em Nature Physics}, 9:200, 2013.

\bibitem{traversa14b}
F.L. Traversa and M.~Di~Ventra.
\newblock Universal memcomputing machines.
\newblock {\em IEEE Trans. Neural Netw. Learn. Syst., (DOI:
  10.1109/TNNLS.2015.2391182, preprint arXiv:1405.0931)}, 2015.

\bibitem{pershin10c}
Yuriy~V. Pershin and Massimiliano Di~Ventra.
\newblock Experimental demonstration of associative memory with memristive
  neural networks.
\newblock {\em Neural {N}etworks}, 23:881, 2010.

\bibitem{chua76a}
Leon~O. Chua and Sung~Mo Kang.
\newblock Memristive devices and systems.
\newblock {\em Proc. IEEE}, 64:209--223, 1976.

\bibitem{diventra09a}
Massimiliano {Di Ventra}, Yuriy~V. Pershin, and Leon~O. Chua.
\newblock Circuit elements with memory: Memristors, memcapacitors, and
  meminductors.
\newblock {\em Proc. IEEE}, 97(10):1717--1724, 2009.

\bibitem{pershin09b}
Y.~V. Pershin, S.~{La Fontaine}, and M.~{Di Ventra}.
\newblock Memristive model of amoeba learning.
\newblock {\em Phys. Rev. E}, 80:021926, 2009.

\bibitem{traversa13a}
F.~L. Traversa, Y.~V. Pershin, and M.~Di~Ventra.
\newblock Memory models of adaptive behaviour.
\newblock {\em IEEE Trans. Neural Netw. Learn. Syst.}, 24:1437 -- 1448, 2013.

\bibitem{jo10a}
Sung~Hyun Jo, Ting Chang, Idongesit Ebong, Bhavitavya~B. Bhadviya, Pinaki
  Mazumder, and Wei Lu.
\newblock Nanoscale memristor device as synapse in neuromorphic systems.
\newblock {\em Nano Lett.}, 10:1297--1301, 2010.

\bibitem{Kim12}
Kuk-Hwan Kim, Siddharth Gaba, Dana Wheeler, Jose~M. Cruz-Albrecht, Tahir
  Hussain, Narayan Srinivasa, and Wei Lu.
\newblock A functional hybrid memristor crossbar-array/cmos system for data
  storage and neuromorphic applications.
\newblock {\em Nano Letters}, 12(1):389--395, 2012.

\bibitem{pershin12a}
Y.~V. Pershin and M.~{Di Ventra}.
\newblock Neuromorphic, digital and quantum computation with memory circuit
  elements.
\newblock {\em Proc. {IEEE}}, 100:2071, 2012.

\bibitem{borghetti10a}
Julien Borghetti, Gregory~S. Snider, Philip~J. Kuekes, J.~Joshua Yang,
  Duncan~R. Stewart, and R.~Stanley Williams.
\newblock {`Memristive' switches enable `stateful' logic operations via
  material implication}.
\newblock {\em Nature}, {464}:{873--876}, {2010}.

\bibitem{pershin11d}
Yuriy~V. Pershin and Massimiliano Di~Ventra.
\newblock Solving mazes with memristors: a massively-parallel approach.
\newblock {\em Phys. Rev. E}, 84:046703, 2011.

\bibitem{xia09a}
Qiangfei Xia, Warren Robinett, Michael~W. Cumbie, Neel Banerjee, Thomas~J.
  Cardinali, J.~Joshua Yang, Wei Wu, Xuema Li, William~M. Tong, Dmitri~B.
  Strukov, Gregory~S. Snider, Gilberto Medeiros-Ribeiro, and R.~Stanley
  Williams.
\newblock Memristor-{CMOS} hybrid integrated circuits for reconfigurable logic.
\newblock {\em Nano Letters}, 9:3640--3645, 2009.

\bibitem{pershin13b}
Yuriy~V. Pershin and Massimiliano Di~Ventra.
\newblock Self-organization and solution of shortest-path optimization problems
  with memristive networks.
\newblock {\em Phys. Rev. E}, 88:013305, Jul 2013.

\bibitem{Wright11a}
C.~David Wright, Yanwei Liu, Krisztian~I. Kohary, Mustafa~M. Aziz, and
  Robert~J. Hicken.
\newblock Arithmetic and biologically-inspired computing using phase-change
  materials.
\newblock {\em Advanced Materials}, 23:3408--3413, 2011.

\bibitem{thomas2013memristor}
Andy Thomas.
\newblock Memristor-based neural networks.
\newblock {\em Journal of Physics D: Applied Physics}, 46(9):093001, 2013.

\bibitem{pershin14a}
Yuriy~V. Pershin and Massimiliano Di~Ventra.
\newblock Memcapacitive neural networks.
\newblock {\em Electronics Letters}, 50:141, 2014.

\bibitem{linn2012beyond}
E~Linn, R~Rosezin, S~Tappertzhofen, U~B{\"o}ttger, and R~Waser.
\newblock Beyond von {N}eumann-logic operations in passive crossbar arrays
  alongside memory operations.
\newblock {\em Nanotechnology}, 23(30):305205, 2012.

\bibitem{ievlev14a}
A.~V. Ievlev, S.~Jesse, A.~N. Morozovska, E.~Strelcov, E.~A. Eliseev, Y.~V.
  Pershin, A.~Kumar, V.~Ya. Shur, and S.~V. Kalinin.
\newblock Intermittency, quasiperiodicity and chaos in probe-induced
  ferroelectric domain switching.
\newblock {\em Nature Physics}, 10:59--66, 2014.

\bibitem{Backus78a}
J.~Backus.
\newblock Can programming be liberated from the von {N}eumann style? a
  functional style and its algebra of programs.
\newblock {\em Comm. ACM}, 21:613--641, 1978.

\bibitem{traversa14a}
F.~L. Traversa, F.~Bonani, Y.~V. Pershin, and M.~Di~Ventra.
\newblock Dynamic computing random access memory.
\newblock {\em Nanotechnology}, 25:285201, 2014.

\bibitem{Kogge11a}
P.~Kogge.
\newblock The tops in flops.
\newblock {\em {IEEE Spectrum}}, 48:48--54, 2011.

\bibitem{martinez09a}
J.~Martinez-Rincon, Massimiliano Di~Ventra, and Yuriy~V. Pershin.
\newblock Solid-state memcapacitive system with negative and diverging
  capacitance.
\newblock {\em Phys. Rev. B}, 81:195430, 2010.

\bibitem{pershin11c}
J.~Martinez-Rincon and Yuriy~V. Pershin.
\newblock Bistable non-volatile elastic membrane memcapacitor exhibiting
  chaotic behavior.
\newblock {\em IEEE Trans. El. Dev.}, 58:1809, 2011.

\bibitem{Liu06a}
Shangqing Liu, Naijuan Wu, Alex Ignatiev, and Jianren Li.
\newblock Electric-pulse-induced capacitance change effect in perovskite oxide
  thin films.
\newblock {\em J. Appl. Phys.}, 100:056101, 2006.

\bibitem{Lai09a}
Qianxi Lai, Lei Zhang, Zhiyong Li, William~F. Stickle, R.~Stanley Williams, and
  Yong Chen.
\newblock Analog memory capacitor based on field-configurable ion-doped
  polymers.
\newblock {\em Appl. {P}hys. {L}ett.}, 95:213503, 2009.

\bibitem{Flak14a}
J~Flak, E~Lehtonen, M~Laiho, A~Rantala, M~Prunnila, and T~Haatainen.
\newblock Solid-state memcapacitive device based on memristive switch.
\newblock {\em Semiconductor Science and Technology}, 29(10):104012, 2014.

\bibitem{eichler11}
A.~Eichler, Joel Moser, J.~Chaste, M.~Zdrojek, I.~Wilson-Rae, and Adrian
  Bachtold.
\newblock Nonlinear damping in mechanical resonators made from carbon nanotubes
  and graphene.
\newblock {\em Nature nanotechnology}, 6(6):339--342, 2011.

\bibitem{stoller08}
Meryl~D. Stoller, Sungjin Park, Yanwu Zhu, Jinho An, and Rodney~S. Ruoff.
\newblock Graphene-based ultracapacitors.
\newblock {\em Nano Letters}, 8(10):3498--3502, 2008.

\bibitem{ElKadi13}
Maher~F. El-Kady and Richard~B. Kaner.
\newblock Scalable fabrication of high-power graphene micro-supercapacitors for
  flexible and on-chip energy storage.
\newblock {\em Nature Communications}, 4:1475, 2013.

\bibitem{Singh14a}
V.~Singh, S.~J. Bosman, B.~H. Schneider, Y.~M. Blanter, Castellanos-Gomez A.,
  and G.~A. Steele.
\newblock Optomechanical coupling between a multilayer graphene mechanical
  resonator and a superconducting microwave cavity.
\newblock {\em Nature Nanotechnology (advanced online publication)}, 2014.

\bibitem{Biolek10b}
D.~Biolek, Z.~Biolek, and V.~Biolkova.
\newblock {SPICE} modelling of memcapacitor.
\newblock {\em El. Lett.}, 46:520, 2010.

\bibitem{krems2010a}
Matt Krems, Yuriy~V. Pershin, and Massimiliano Di~Ventra.
\newblock Ionic memcapacitive effects in nanopores.
\newblock {\em Nano Lett.}, 10:2674, 2010.

\bibitem{pershin11a}
Yuriy~V. Pershin and Massimiliano Di~Ventra.
\newblock Memory effects in complex materials and nanoscale systems.
\newblock {\em Advances in Physics}, 60:145--227, 2011.

\bibitem{graphene2}
Sukanta De and Jonathan~N. Coleman.
\newblock Are there fundamental limitations on the sheet resistance and
  transmittance of thin graphene films?
\newblock {\em ACS Nano}, 4(5):2713--2720, 2010.
\newblock PMID: 20384321.

\bibitem{graphene3}
Ki~Kang Kim, Alfonso Reina, Yumeng Shi, Hyesung Park, Lain-Jong Li, Young~Hee
  Lee, and Jing Kong.
\newblock Enhancing the conductivity of transparent graphene films via doping.
\newblock {\em Nanotechnology}, 21(28):285205, 2010.

\bibitem{graphene4}
Hongtao Liu, Yunqi Liu, and Daoben Zhu.
\newblock Chemical doping of graphene.
\newblock {\em J. Mater. Chem.}, 21:3335--3345, 2011.

\bibitem{petkov04a}
V.P. Petkov and B.E. Boser.
\newblock Capacitive interfaces for mems.
\newblock In H.~Baltes, O.~Brand, G.~K. Fedder, C.~Hierold, J.~G. Korvink, and
  O.~Tabata, editors, {\em Advanced Micro and Nanosystems}, pages 49--92.
  Wiley-VCH, Weinheim, Weinheim, 2004.

\bibitem{strukov05a}
DB~Strukov and KK~Likharev.
\newblock {CMOL FPGA: a reconfigurable architecture for hybrid digital circuits
  with two-terminal nanodevices}.
\newblock {\em {Nanotechn.}}, {16}:{888--900}, {2005}.

\end{thebibliography}
\bibliographystyle{unsrt}



\end{document}